\documentclass[a4paper,11pt]{article}
\usepackage{jcappub} 
\usepackage{amsfonts}
\usepackage{caption}
\usepackage{subcaption}
\usepackage{float}
\usepackage{mathtools}

\title{\boldmath Phase space analysis of $f(R)$ cosmologies: revisiting attractor-to-GR}

\author[1]{Gahan Chattopadhyay\note{gahanch080@gmail.com}}
\author[2]{and Soumitra Sengupta\note{tpssg@iacs.res.in}}
\affiliation{School of Physical Sciences, Indian Association for the Cultivation of Science,\\2A $\&$ 2B, Raja S.C. Mullick Road Kolkata - 700 032, India}

\abstract{Phase space analysis of $f(R)$-Gravity, a simple generalization of Einstein's General theory of Relativity has been of interest especially in the context of Cosmology. Beyond its success in providing some viable descriptions of both early and late time cosmology, $f(R)$ gravity is considered to be an ideal set up for investigating some generic features of modified gravitational dynamics because of its mathematical simplicity, freedom from higher-derivative ghost instabilities and also having a natural dual description as scalar-tensor gravity in the Einstein frame representation. In this work, we have considered a class of $f(R)$-Gravity theories with $f(R)=R+\alpha R^n$ for both positive and negative integral values of $n$ and studied some general aspects of their cosmological phase space dynamics in the Einstein frame. We have used standard techniques from dynamical systems to obtain finite fixed points and assess their stability non-perturbatively by the Lyapunov method, whereas fixed points at infinity have been addressed using Poincar\'e projection. We have introduced techniques from Hybrid dynamical systems to the models with odd $n$, which do not admit standard methods of analysis in the Einstein frame representation due to peculiarities of the conformal transformation involved. Based on both analytical and numerical results we show the claimed convergence of all such $f(R)$ theories to GR (plus a dark energy like component for inverse power models) in a universe with a future-singularity free expansion history, assuming an isotropic and homogeneous background - via the dynamical freeze-out of the extra scalar degree of freedom to stable fixed points. At the end we discuss some observational constraints on the parameters of such theories and discuss their consistency with respect to our work.}


\begin{document}
\maketitle
\flushbottom

\section{Introduction}
General Relativity provides an exceptionally successful description of
gravitational phenomena over a wide range of scales. Nevertheless, extensions
of the Einstein--Hilbert action by nonlinear functions of the Ricci scalar have
long been investigated as a possible description of gravitational dynamics
beyond General Relativity. In particular, 
the class of gravitational theories that arises from the simple generalization of the Einstein-Hilbert action from $R$ to some arbitrary function $f(R)$ of the Ricci scalar $R$, is called $f(R)$-gravity \cite{Sotiriou:2008rp, Nojiri:2003ft, Nojiri:2006ri, Nojiri:2017ncd}. Such theories were also previously shown to have dual scalar-tensor theories, related by a conformal transformation of the metric tensor, where the scalar sector encodes the information about the higher derivative corrections present in the corresponding $f(R)$-theory \cite{Nojiri:2010wj, Maeda:1988ab, Magnano:1993bd, Fujii:2007qv, Armendariz-Picon:2000nqq}. This scalar degree of
freedom also provides a useful way of studying the cosmological dynamics of
the higher-derivative corrections as a nonlinear dynamical system. The function $f(R)$ can contain terms such as $R^2$, $R^3$ and higher positive powers of $R$ that dominate the leading order term $R$ only in strong curvature regimes and also may have terms such as $\frac{1}{R}$, $\frac{1}{R^2}$ and higher negative powers of $R$ that dominate $R$ in low curvature regimes. 

\noindent Positive power corrections $R^n(n\geq2)$ have been studied extensively before \cite{DeFelice:2010aj, Capozziello:2002rd, Amendola:2006we, Whitt:1984pd, Cembranos:2008gj, Hu:2007nk, Schmidt:1990gb}, one of the most well-known of which is the Starobinsky theory of Gravity with $f(R)=R + \alpha R^2$, which incorporates inflation in the early universe without introducing an Inflaton field by hand \cite{Starobinsky:1980te, Vilenkin:1985md, Armendariz-Picon:2000ulo, Capozziello:2003tk, Odintsov:2025zrp}. But since the positive-power modifications are suppressed by higher powers of the Planck mass ($M_p$), their effects essentially wash out due to cosmological evolution of the universe leading to lower values of the curvature $R$ and thus giving Einstein-gravity as a low energy effective theory for the gravitational sector in the late Universe. Negative or inverse power corrections $R^n(n < 0)$ have also been explored in \cite{Nojiri:2003ft, Carroll:2003wy} to explain the late-time accelerated expansion of our universe. However, later these models were found to suffer from the infamous Dolgov-Kawasaki Instability \cite{Dolgov:2003px, Capozziello:2003gx}, which questioned the viability of such models. In a slightly different context but more interesting and recent work, the potential of non-minimally coupled (to matter) $f(R)$ models, with inverse powers of $R$, to solve the Hubble tension was explored in \cite{BarrosoVarela:2024htf}, and statistical analysis of that model on DESI/DES/Pantheon+/eBOSS data showed significant evidence towards its preference over $\Lambda$-CDM cosmology \cite{BarrosoVarela:2024ozs}.

\noindent The cosmological dynamics of $f(R)$ theories have been studied extensively
using dynamical-systems methods. In particular, expansion-normalized
formulations have been developed for general $f(R)$ theories and applied to
power-law models, including $R+\alpha R^n$ gravity
\cite{Carloni:2015jla}. Related investigations of scalar-tensor cosmologies have
studied the phase-space structure close to the general-relativity limit and
the conditions under which this limit can act as an attractor
\cite{Jarv:2010zc, Jarv:2010xm, Jarv:2015kga}. This attractor-to-GR mechanism has also been shown to work in the matter-dominated scenario and in the setting of Teleparallel gravity as well \cite{Jarv:2011sm, Jarv:2015odu}. More generally, the phase-space
geometry of scalar-field cosmology, including the Hamiltonian constraint
surface, finite equilibrium points, asymptotic configurations and
non-perturbative stability through Lyapunov functions, has been systematically
studied by Faraoni and Protheroe \cite{Faraoni:2012bf}. These works
provide both intuition as well as important tools for understanding the dynamical fate of the extra gravitational degrees
of freedom in modified-gravity cosmologies.

\noindent In the present work, we revisit the global phase-space cosmological dynamics of the extra scalar degree of freedom associated with the family $f(R)=R +\alpha R^n$ for all integral values of $n$. Our goal is not simply to identify isolated fixed points, but to characterize the finite and asymptotic structures of the phase space in the Einstein frame representation. The finite fixed points are obtained from local conditions on the full cosmological equations whereas the asymptotic structure is investigated employing Poincar\'e compactification. The resulting phase portraits are then compared with direct numerical evolution. This provides a unified description of the dynamical fate of the extra scalar degree of freedom across the different members of the $R+\alpha R^n$ family.
\noindent A further subtlety arises for odd values of $n$. The conformal transformation relating the Jordan-frame curvature to the Einstein-frame scalar is not globally one-to-one in these cases: positive and negative values of $R$ can be mapped to the same value of the Einstein-frame scalar. In particular, for positive odd $n$, the Jordan-frame theory is regular at $R=0$, and a Jordan-frame trajectory can pass continuously from $R>0$ to $R<0$. The same evolution, however, reaches the boundary $\sigma=0$ in the Einstein frame, where the two Jordan-frame branches are identified. The ordinary Einstein-frame potential does not by itself encode the continuation across this boundary. Consequently, a global description in terms of a single
smooth Einstein-frame vector field is insufficient. We formulate this sector using the language of Hybrid dynamical systems \cite{e4de97e1c1334dc99ffa1394e35ae6d9}, specifying the continuous flow, the event time and the corresponding impact rule required to represent the underlying Jordan-frame evolution consistently in the Einstein-frame phase space. \\

\noindent The phase-space analysis is also extended to cosmological backgrounds containing ordinary matter. We consider both radiation and dust and examine the resulting fixed-point structure and stability, thereby determining how the presence of matter affects the vacuum phase-space dynamics and the approach towards the general-relativity regime. Finally, we discuss the phenomenological constraints on the model parameters. In particular, we compare the parameter ranges considered in the dynamical analysis with constraints from inflationary observables and late-time cosmology, including those relevant to viable inverse-power $f(R)$ models. \\

\noindent The paper is organized as follows. In section [\ref{sec:review}] we briefly review $f(R)$ gravity and its Einstein-frame scalar-tensor representation. Section [\ref{sec:setup}] introduces the $R+\alpha R^n$ models and derives the Einstein-frame potential and cosmological equations. In section [\ref{sec:dynam}] we construct the physical phase space, identify the finite and asymptotic fixed points, and analyze their stability, treating positive even, positive odd and inverse-power models separately. The hybrid formulation required for the odd-power sector is developed there as well. Section [\ref{sec:matter}] extends the analysis to radiation and dust. Section [\ref{sec:observations}] discusses the observational constraints on the model parameters and their relation to the dynamical results. Finally, section [\ref{sec:conclusion}] summarizes our results and discusses possible extensions of the analysis.

\section{Brief review of f(R) theory and dual Scalar-Tensor representation} \label{sec:review}
The primary motivation behind considering f(R) theories of gravity comes from loop corrections to matter fields in curved space-time \cite{Utiyama:1962sn,Franchino-Vinas:2019upg} and their potential in explaining early-time inflation\cite{Odintsov:2022bpg, Odintsov:2022rok, Odintsov:2020thl} and late-time acceleration of the universe\cite{Nojiri:2008nt, Cognola:2007zu, Capozziello:2006dj, Nojiri:2007cq, Nojiri:2009kx, Odintsov:2025jfq, Odintsov:2023cli}. The most general diffeomorphism-invariant f(R) action that one can write is:
\begin{equation}\label{eq:fRaction}
    S=M_p^2\int d^4x\sqrt{-g}f(R)
\end{equation} 
where 
\begin{equation}\label{eq:fR}
    f(R)=\sum_{i=-\infty}^{\infty}\alpha_iR^i
\end{equation} 
contains the EH term $(i=1)$ along with both positive and inverse power corrections. The coefficients $\alpha_i$ are of appropriate dimensions and scaling factors as suggested by effective field theory. Now, the action (\ref{eq:fRaction}) can be written in a dynamically equivalent form by introducing an auxiliary field $\chi$ in the following way:
\begin{equation}\label{eq:jordan}
    S=M_p^2\int d^4x\sqrt{-g}\left(f'(\chi)(R-\chi) + f(\chi)\right)
\end{equation}
One can verify that variation with respect to $\chi$ gives the constraint equation $f''(\chi)(R-\chi)=0$ which implies that $R=\chi$ because $f''(\chi)\neq 0$ due to nonlinear $f$ and thus get back action (\ref{eq:fRaction}). In literature the action (\ref{eq:jordan}) is called the Jordan frame action. Now consider the conformal transformation of the form $g_{\mu\nu}\xrightarrow{}e^{-\sigma/\sqrt{3}M_p}g_{\mu\nu}$ where $\sigma/\sqrt{3}M_p = -ln|f'(\chi)|$. If one rewrites the action in terms of the transformed metric and the scalar field $\sigma$,
one notices that the action becomes that of Einstein gravity and a minimally coupled scalar field with self-interaction potential $V(\sigma)$: 
\begin{align}
    &S = M_p^2\int d^4x\sqrt{-g} R-\int d^4x\sqrt{-g}\left(\frac{1}{2}g^{\mu\nu}\partial_\mu \sigma\partial_\nu\sigma+V(\sigma)\right) \label{eq:Eaction} \\
    \text{where }&V(\sigma)=V(R(\sigma))=M_p^2\frac{Rf'(R)-f(R)}{f'(R)^2} \label{eq:Vint}
\end{align}

\noindent In the $f(R)$ action, the condition $f'(R)>0$ is essential so that the corresponding scalar field has the correct sign before its kinetic term \cite{deRham:2016wji, Woodard:2015zca}. The action (\ref{eq:Eaction}) is called the Einstein frame action. Here, onward, we will only work with the Einstein frame action and refer only to the corresponding choices of $f(R)$. This scalar field in the dual theory is essentially the higher-curvature degree of freedom from the original $f(R)$ theory. For all integral values of $n$ in $f(R)=R+\alpha R^n$ we will specifically look at solutions which correspond to an expanding universe in the Jordan frame. However, we will formulate the dynamical system in the corresponding Einstein frame representation and study the cosmological phase space dynamics both analytically and numerically, explicitly showing the existence of a stable GR fixed point.

\section{The general setup} \label{sec:setup}
\subsection{Einstein frame representation of $f(R)=R+\alpha R^n$}

\begin{figure}[!h]
    \centering

    \begin{subfigure}{0.45\linewidth}
        \centering
        \includegraphics[width=0.8\linewidth]{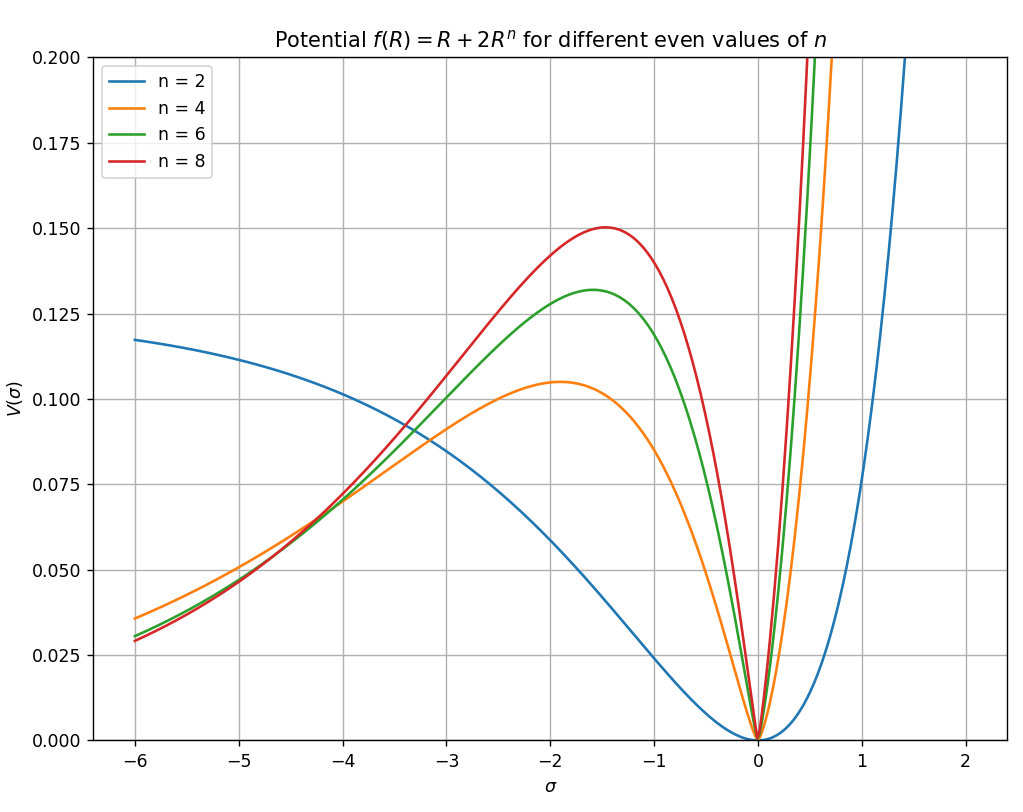}
        \caption{Varying even $n$ for $\alpha=2M_p^{2-2n}$}\label{fig:a}
    \end{subfigure}
    \hspace{0.3cm}
    \begin{subfigure}{0.45\linewidth}
        \centering
        \includegraphics[width=0.8\linewidth]{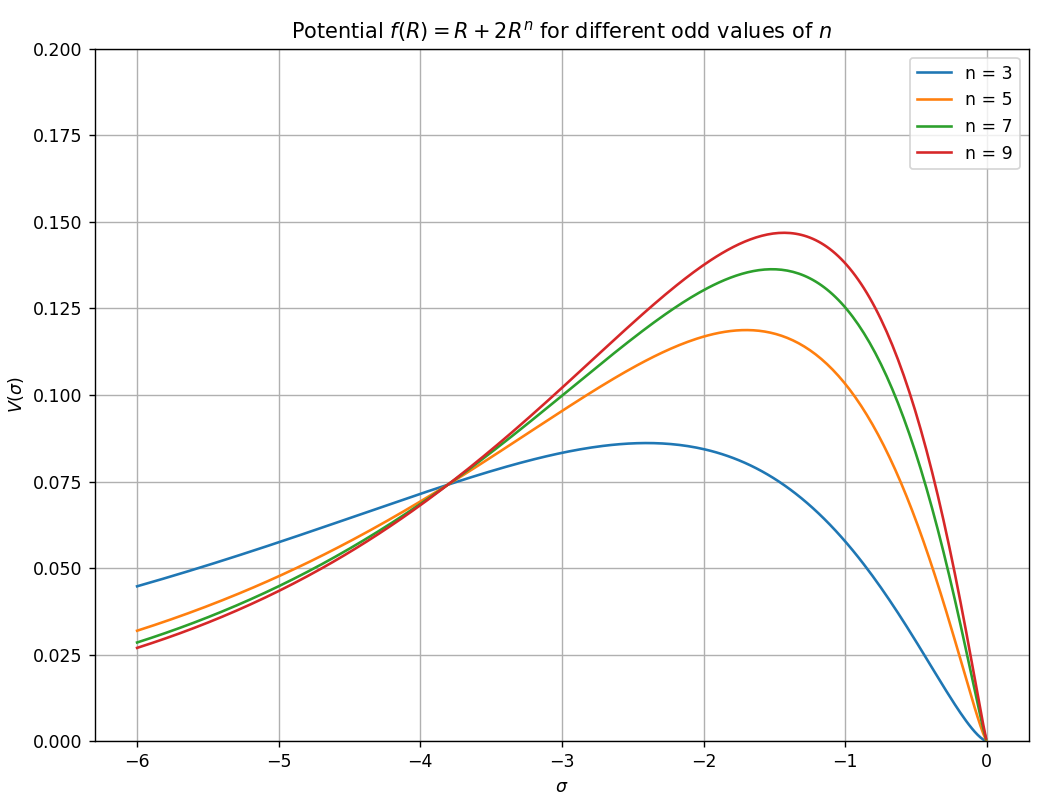}
        \caption{Varying odd $n$ for $\alpha=2M_p^{2-2n}$}\label{fig:b}
    \end{subfigure}

    \vspace{0.5cm}

    \begin{subfigure}{0.45\linewidth}
        \centering
        \includegraphics[width=0.8\linewidth]{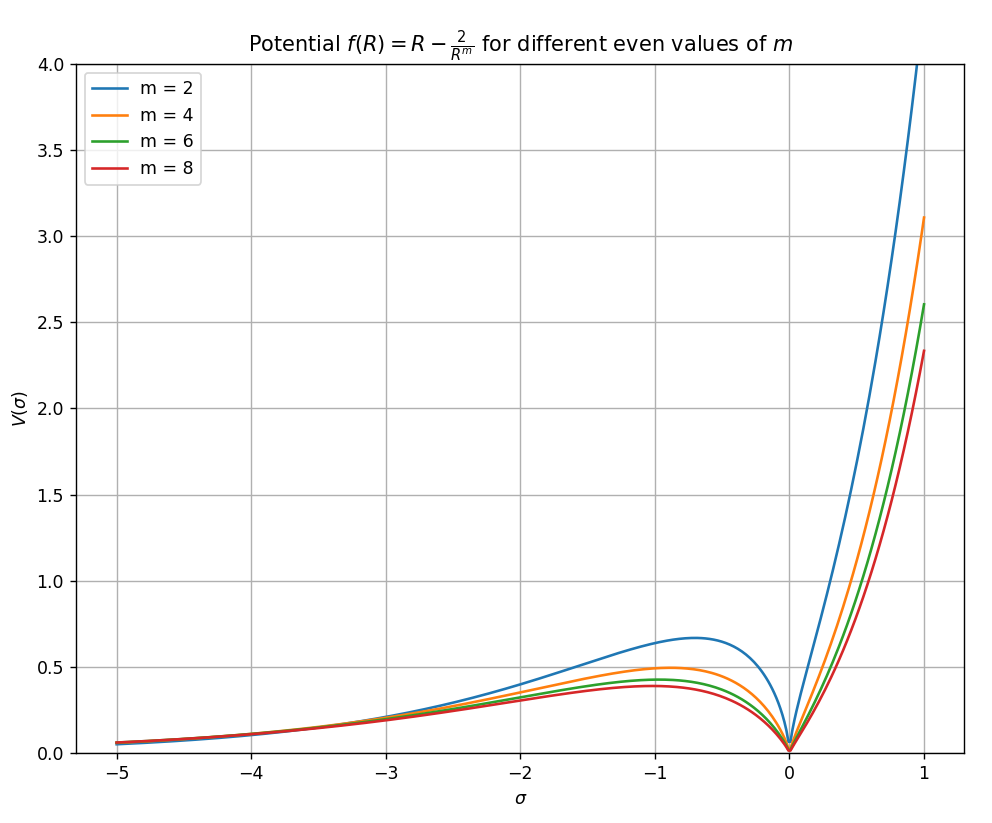}
        \caption{Varying even $n=-m$ for $\alpha=-2M_p^{2+2m}$}\label{fig:c}
    \end{subfigure}
    \hspace{0.3cm}
    \begin{subfigure}{0.45\linewidth}
        \centering
        \includegraphics[width=0.8\linewidth]{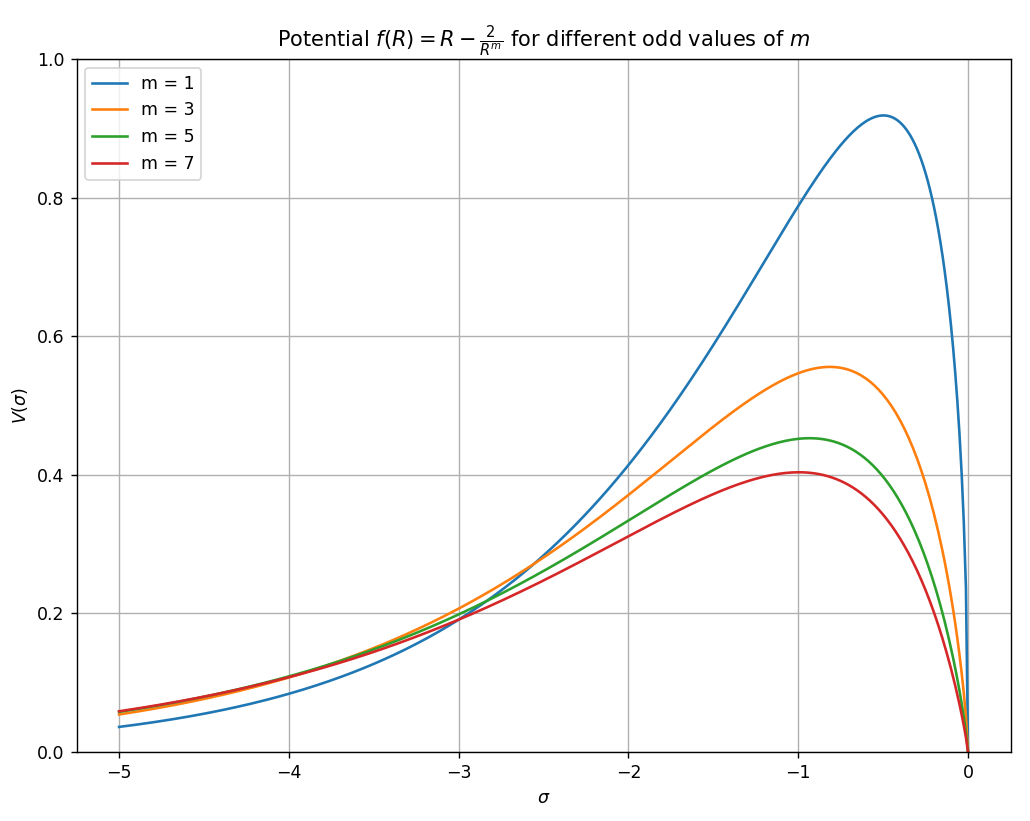}
        \caption{Varying odd $n=-m$ for $\alpha=-2M_p^{2+2m}$}\label{fig:d}
    \end{subfigure}

    \caption{Forms of the Einstein frame potentials for various choices of positive $n$ as well as negative ($n=-m$). Here, $\sigma$ is in units of $M_p$.}
    \label{fig:panel1}
\end{figure}

Determining the Einstein frame representation of the $f(R)$ theory simply amounts to finding the $R$-to-$\sigma$ map and calculating the potential $V(\sigma)$. To calculate the corresponding potential in the Einstein frame, we will use the relation $\sigma/(\sqrt{3}M_p)=-ln\left[f'(\chi)\right]$ and invert to find $\chi(\sigma)$ first.
\begin{align}
   & f'(\chi)=e^{-\sigma/\sqrt{3}M_p}=1+\alpha n\chi^{n-1} \\
   & \implies \chi(\sigma)=R(\sigma)=\left(\frac{e^{-\sigma/\sqrt{3}M_p}-1}{\alpha n}\right)^{\frac{1}{n-1}} \label{eq:RmapSigma}
\end{align}
Then, we can obtain an explicit form of the potential $V(\sigma)$ by substituting $R(\sigma)$ into equation (\ref{eq:Vint}),
\begin{align}
    \begin{split}
        V(\sigma)&=M_p^2\frac{R(\sigma)(1+\alpha nR(\sigma)^{n-1})-(R(\sigma)+\alpha R(\sigma)^n)}{e^{-2\sigma/\sqrt{3}M_p}} \\ 
        &=M_p^2\alpha(n-1)R(\sigma)^ne^{2\sigma/\sqrt{3}M_p} \\
        &=M_p^2\alpha(n-1)\left(\frac{e^{-\sigma/\sqrt{3}M_p}-1}{\alpha n}\right)^{\frac{n}{n-1}}e^{2\sigma/\sqrt{3}M_p} \label{eq:Vsigma}
    \end{split}
\end{align}
At this point it is crucial to note that for odd values of $n$ (either positive or negative) the map $R\mapsto\sigma$ is not one-to-one (see equation \ref{eq:RmapSigma}). Both positive and negative $R$ are mapped to the same $\sigma$ which is negative. As a consequence, the potential in these cases will only be real on the negative $\sigma$ side of the axis, as can be seen from figures [\ref{fig:panel1}]. For the negative odd values $\sigma\rightarrow0$ implies $R\rightarrow\infty$ which is a curvature singularity - this will reflect in the derivative of the potential near $\sigma=0$ as we will see later in section [\ref{sec:inverse}]. However for the positive odd values $\sigma\rightarrow0$ implies $R\rightarrow0$, meaning there is no singularity. Therefore, during dynamics when $R$ flows smoothly from the positive to the negative values, $\sigma$ essentially must bounce back from the point $\sigma=0$ in order to be a consistent representation of the Jordan frame system. But the potential provides no such reflecting condition on its own, it has to be imposed by hand if the dynamical systems analysis has to be performed in the Einstein frame representation. This will turn out to be the crucial difference in the phase space analysis for the case of odd values of $n$ and will require specialized techniques [\ref{sec:hybrid}].

\subsection{Cosmological evolution of the scalar degree of freedom}
We will primarily focus on vacuum evolution, i.e - solutions without matter fields or with sub-dominant contribution from matter fields. Inclusion of matter complicates the phase dynamics of the theory considerably, however some general comments and numerical results in such scenarios will be discussed later in this article. Varying the action 
(\ref{eq:Eaction}) with respect to the metric gives the gravitational field equations. In a flat isotropic and homogeneous space-time,
\begin{equation}
    ds^2=-dt^2+a(t)^2dx_idx^i
\end{equation}
these boil down to two equations (the Friedmann equation and the acceleration equation). On the other hand, varying the action with respect to the scalar field $\sigma$ gives its equation of motion. Of these three equations, only two are independent. We will choose to work with the Friedmann equation and the equation of motion of $\sigma$ : 
\begin{align}
    & 3H^2=\frac{1}{2M_p^2}\left( \frac{1}{2}\dot{\sigma}^2 + V(\sigma) \right) \label{eq:hamiltonC} \\
    & \ddot{\sigma}+3H\dot{\sigma}+V'(\sigma)=0 \label{eq:eomSc}
\end{align}

\section{The dynamical system: Analysis of fixed points and their stability} \label{sec:dynam}

While formulating the dynamical system for maximally symmetric scalar field cosmologies, in general the phase space is taken to be the four dimensional space $(a,\dot a, \sigma, \dot\sigma)$ along with a first order Hamiltonian constraint which reduces the dynamics of the system to orbits on a 3 dimensional hypersurface. But, in the spatially flat $(k=0)$ case, an additional simplification happens: the scale factor only appears as the combination $(\frac{\dot a}{a})\equiv H$. Therefore the dynamical variables become $(H, \sigma,\dot \sigma)$. On the other hand the Hamiltonian constraint (\ref{eq:hamiltonC}) renders one of the three variables dependent and reduces the dynamics to orbits on a 2 dimensional surface. One can thereafter choose to work with any two of the three variables.

\subsection{Phase space geometry and Finite fixed points}\label{sec:psg}

\begin{figure}[!h]
    \centering

    \begin{subfigure}{0.45\linewidth}
        \centering
        \includegraphics[width=0.8\linewidth]{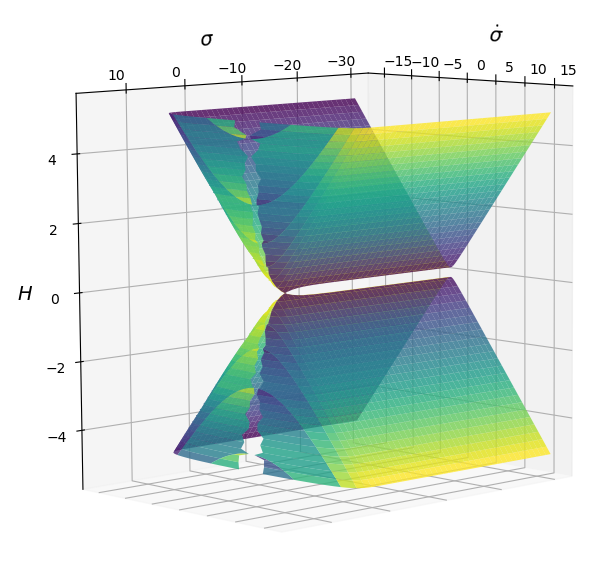} 
        \caption{$n=2$, $\alpha=2M_p^{2-2n}$}\label{fig:a}
    \end{subfigure}
    \vspace{0.5cm}
    \begin{subfigure}{0.45\linewidth}
        \centering
        \includegraphics[width=0.8\linewidth]{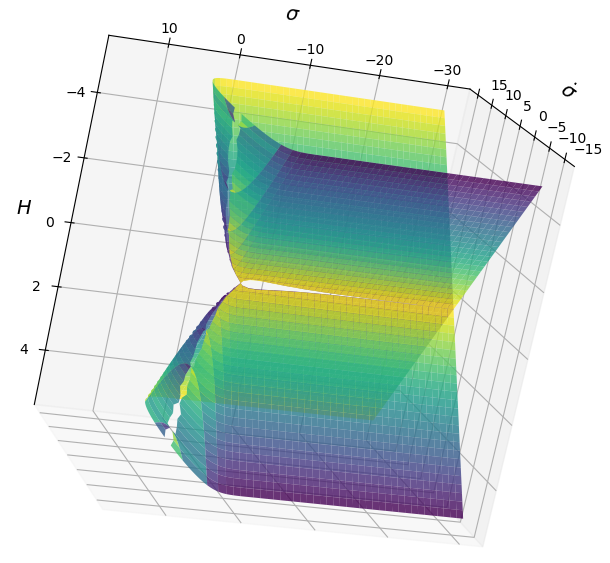} 
        \caption{$n=4$, $\alpha=2M_p^{2-2n}$}\label{fig:b}
    \end{subfigure}
    \hspace{0.3cm}
    \begin{subfigure}{0.45\linewidth}
        \centering
        \includegraphics[width=0.8\linewidth]{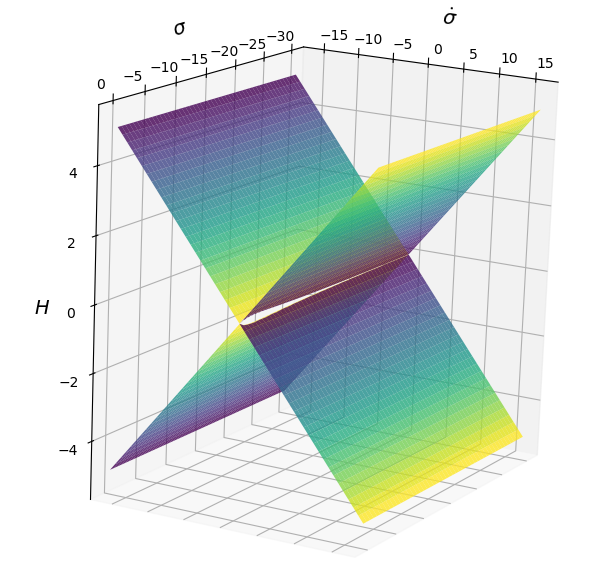}
        \caption{$n=3$, $\alpha=2M_p^{2-2n}$}\label{fig:c}
    \end{subfigure}

    \vspace{0.5cm}

    \begin{subfigure}{0.45\linewidth}
        \centering
        \includegraphics[width=0.8\linewidth]{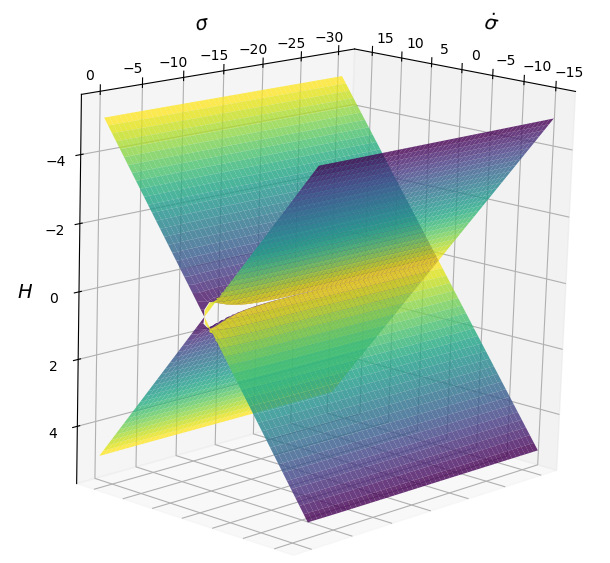}
        \caption{$n=-m=-1$, $\alpha=-2M_p^{2+2m}$}\label{fig:d}
    \end{subfigure}
    \hspace{0.3cm}
    \begin{subfigure}{0.45\linewidth}
        \centering
        \includegraphics[width=0.8\linewidth]{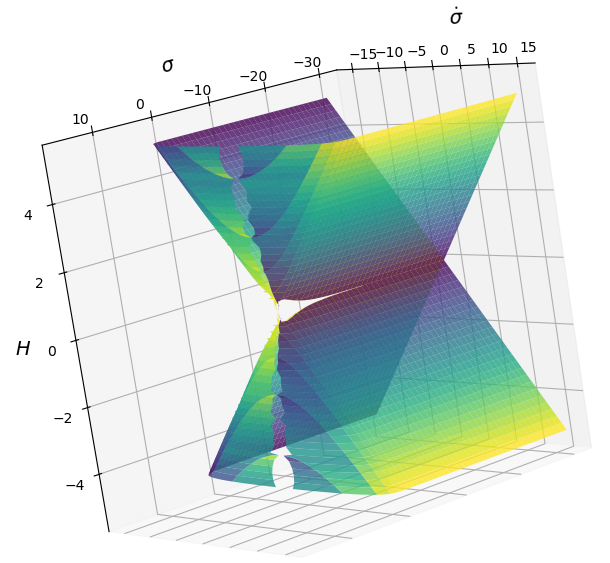}
        \caption{$n=-m=-2$, $\alpha=-2M_p^{2+2m}$}\label{fig:e}
    \end{subfigure}

    \caption{Hamiltonian Constraint surface for a few choices of parameters $n$ and $\alpha$.}
     \label{fig:HCs}
\end{figure}

If one chooses to work with $(H,\sigma)$ then, the 2D surface of constraint will be given by:
\begin{equation}
    \dot\sigma=\pm \sqrt{12M_p^2H^2-2V(\sigma)} \label{eq:constr}
\end{equation}
The two signs on the RHS of equation (\ref{eq:constr}) label the `positive' and `negative' sheets of the constraint surface. The sheets are either disconnected or connected only at the $\dot\sigma=0$ plane, about which they are symmetric. The constraint surface has been plotted for a few parameter choices in Figures \ref{fig:HCs}. 

\noindent The fixed points of this dynamical system are, by definition, of the form $(\dot H,\dot\sigma)\equiv(0,0)$ along with $(\ddot H,\ddot\sigma)\equiv(0,0)$ or simply $(H,\sigma)\equiv(H_0,\sigma_0)$ where $H_0$ and $\sigma_0$ are constants. Therefore the fixed points are de Sitter spaces with a constant scalar profile. The case $H_0=0$ corresponds to Minkowski space. Such fixed points can be identified by solving the following algebraic equations:
\begin{align}
    & H_0^2=\frac{V(\sigma_0)}{6M_p^2}, \,\,\,\,\,\,\,\,V(\sigma_0)\geq 0 \label{eq:ffp1} \\
    & V'(\sigma_0)=0 \label{eq:ffp2}
\end{align}
It should also be noted that in our models, the finite fixed points can also be visually identified from the corresponding plot of the constraint surface. For example, for $n=2$ the positive and negative sheets meet only once at the $\dot \sigma=0$ plane at which point $(H_0,\sigma_0)\equiv(0,0)$, which is a global attractor as will be shown by explicit calculations in section [\ref{sec:poseven}]. Similarly, for $n=3$ and $n=4$ the two sheets meet at the $\dot\sigma=0$ plane twice, once at $(H_0,\sigma_0)\equiv(0,0)$ and again asymptotically as $(H_0,\sigma_0)\rightarrow(0,-\infty)$. Notice that for $n=-1$ and $n=-2$ as well, the sheets meet asymptotically at $\sigma_0\rightarrow-\infty$, hinting that such $f(R)$ models with $n \neq 2$ might generically have fixed points at infinity. All of these statements will be rigorously verified in the following subsections.

\newpage

\subsection{Poincar\'{e} projection: Fixed points at infinity} \label{sec:infinity}

In order to rigorously establish the existence of fixed points at infinity and characterize them, one utilizes the following transformations known as the Poincar\'{e} projection, which map infinity to the circle $\bar r=1$\ -

\begin{align}
    H=r\cos{\theta ,\,\,\,\, \sigma=r\sin{\theta}} \\
    \text{where}\,\,\,\,\,\,\,\, r=\frac{\sqrt{\bar r}}{1-\bar r}
\end{align}

\noindent In terms of the new dynamical variables $(\bar r,\theta)$ the dynamical system given by (\ref{eq:hamiltonC}) and (\ref{eq:eomSc}) becomes -  

\begin{eqnarray}
&&\frac{\dot{\bar{r}}\left(1+\bar{r}\right)}{2\sqrt{\bar{r}} 
\left(1-\bar{r}\right)^{2}}\:\cos\theta 
- \, \frac{\sqrt{\bar{r}}}{1-\bar{r}} 
\:\dot{\theta}\sin\theta \nonumber\\
&&\nonumber\\
&&=  -\frac{1}{4M_p^2}\left[\frac{\dot{\bar{r}} 
\left(1+\bar{r}\right)}{2\sqrt{\bar{r}}\left(1-\bar{r} 
\right)^{2}}\:\sin\theta+ \frac{\sqrt{\bar{r}}}{1-\bar{r}} 
\:\dot{\theta}\cos\theta\right]^{2} 
\,,\label{eq:poincareHdot}
\end{eqnarray}
\begin{eqnarray}
&&\left[\frac{\ddot{\bar{r}}+\ddot{\bar{r}}\bar{r}
+\dot{\bar{r}}^{2}}{2\sqrt{\bar{r}}\left(1-\bar{r}\right)^{2}}
- \, \frac{\dot{\bar{r}}^{2}+\dot{\bar{r}}^{2}\bar{r}}{ 
4\bar{r}^{3/2}\left(1-\bar{r}\right)^{2}} 
+ \, \frac{\dot{\bar{r}}^{2} 
+\dot{\bar{r}}^{2}\bar{r}}{4\sqrt{ 
\bar{r}}\left(1-\bar{r}\right)^{3}}\right] 
\sin\theta \nonumber\\
&&\nonumber\\
&& + \, \frac{\dot{\bar{r}}+\dot{\bar{r}}\bar{r}}{\sqrt{ 
\bar{r}}\left(1-\bar{r}\right)^{2}}\:\dot{\theta}\cos\theta 
+ \, \frac{\sqrt{\bar{r}}}{1-\bar{r}}\left(\ddot{\theta} 
\cos\theta-\dot{\theta}^{2}\sin\theta\right) \nonumber\\
&&\nonumber\\
&&=-3 \left[\frac{\bar{r}}{\left(1-\bar{r}\right)^{2}} 
\:\dot{\theta}\cos^{2}\theta+ \, \frac{\dot{\bar{r}} 
\left(1+\bar{r}\right)}{2\left(1-\bar{r}\right)^{3}} 
\:\sin\theta\right]  - V^{'} 
\,.\label{eq:poincareKG}
\end{eqnarray}

\noindent Even though this system of equations look formidable, due to the linear relation between $H,\sigma$ and $\sin\theta, \cos\theta$, the fixed points $(H_0,\sigma_0)$ simply map to $(\bar r_0,\theta_0)$ and thus the necessary conditions for such a fixed point to exist can be found to be - 

\begin{align}
    & \frac{\bar r \cos^2{\theta}}{(1-\bar r)^2}=\frac{1}{6M_p^2}V(\sigma_0) \label{eq:r1}\\
    & V'(\sigma_0)=0 \\
    & \text{where}\,\,\,\,\,\,\,\, \sigma_0\equiv \sigma(\bar r_0,\theta_0) \label{eq:r2}
\end{align}

\noindent The discussion in \cite{Faraoni:2012bf} lists all the three scenarios in which the equation (\ref{eq:r1}) may be satisfied in the limit $\bar r\rightarrow 1$. As it will become clear in the later subsections, among those scenarios, only one will be relevant to us, which is the case when $\cos{\theta}=0$, implying $H_0\rightarrow 0$, $\sigma_0\rightarrow \pm\infty$ and $V(\sigma_0\rightarrow \pm\infty)=0$. For a discussion of the other two scenarios one may refer to \cite{Faraoni:2012bf}.

\subsection{Lyapunov function and Stability} \label{sec:Lyapunov}
The concept of a Lyapunov function is used to characterize a fixed point's stability non-perturbatively unlike methods involving the study of linear perturbations around the fixed point, which can become cumbersome. The application of the Lyapunov method, laid out in \cite{Faraoni:2012bf}, can be summarized in the following two theorems, given a fixed point $(H_0,\sigma_0)$:
\begin{enumerate}
    \item If $V''(\sigma_0)>0$ and $H_0>0$ then the Lyapunov function $L_1$ is such that:\\
    \begin{itemize}
        \item \[L_1(H,\sigma,\dot\sigma)\coloneqq  \frac{1}{2}\dot\sigma^2 + V(\sigma)-V(\sigma_0)>0\] in a domain $\mathcal{D}$ containing $\sigma_0$ except at $(H_0,\sigma_0)$;
        \item \[L_1(H_0,\sigma_0,0)=0;\]
        \item \[\dot L_1=-3H\dot\sigma^2\] is strictly negative in $\mathcal{D}$, except at $(H_0,\sigma_0)$ where $\dot L_1=0$.
    \end{itemize}
    Therefore such a fixed point is stable. If there is a single stable fixed point, its has a globally attracting basin, otherwise the basins are local. 
    \item If $V''(\sigma_0)<0$ and $H_0>0$ then the Lyapunov function $L_2$ is such that:\\
    \begin{itemize}
        \item \[L_2(H,\sigma,\dot\sigma)\coloneqq  -\frac{1}{2}\dot\sigma^2 - V(\sigma)+V(\sigma_0)<0\] in a domain $\mathcal{D}$ containing $\sigma_0$ except at $(H_0,\sigma_0)$;
        \item \[L_2(H_0,\sigma_0,0)=0;\]
        \item \[\dot L_2=3H\dot\sigma^2\] is strictly positive in $\mathcal{D}$, except at $(H_0,\sigma_0)$ where $\dot L_2=0$.
    \end{itemize}
    Such a fixed point is unstable.
\end{enumerate}

\subsection{Positive Power $n$:}\label{sec:positive}
For positive power models the condition $f'(R)>0$ fixes the sign of the constant $\alpha$ also to be positive. Therefore one can now unambiguously evaluate the slope of the potential as:
\begin{equation}\label{eq:pospot}
    V'(\sigma)=\frac{\alpha M_p^2(n-1)}{(\alpha n)^{\frac{n}{n-1}}}\left[ \frac{2e^{\frac{2\sigma}{\sqrt{3}M_p}}}{\sqrt{3}M_p}\left(e^{-\frac{\sigma}{\sqrt{3}M_p}}-1\right)^{\frac{n}{n-1}}-\frac{ne^{\frac{\sigma}{\sqrt{3}M_p}}}{\sqrt{3}M_p(n-1)}\left(e^{-\frac{\sigma}{\sqrt{3}M_p}}-1\right)^{\frac{1}{n-1}} \right]
\end{equation}
According to equation (\ref{eq:ffp2}) for a fixed point $(H_0,\sigma_0)$, the potential $V(\sigma)$ must have a critical point at $\sigma=\sigma_0$. Solving for the critical points of the potential by setting $V'(\sigma_0)$ from equation (\ref{eq:pospot}) to zero we get the following solutions,
\begin{align}
    & \sigma_0=0\quad \forall\, n\geq2 \label{eq:crit1}\\
    & \sigma_0=-\sqrt{3}M_p\ln{\frac{2n-2}{n-2}}\equiv\sigma_u \quad \forall\, n>2 \label{eq:crit2}
\end{align}
To apply the Lyapunov method to assess stability of the fixed points, it is also necessary to know if the critical point is a local minima or maxima. Therefore the 2nd derivative of the potential is also given as-
\begin{align}\label{eq:pospot2}
V''(\sigma)
&= \frac{\alpha M_p^2(n-1)}{(\alpha n)^{\frac{n}{n-1}}}
\Biggl[
\frac{4e^{\frac{2\sigma}{\sqrt{3}M_p}}}{3M_p^2}
\left(e^{-\frac{\sigma}{\sqrt{3}M_p}}-1\right)^{\frac{n}{n-1}}-\frac{ne^{\frac{\sigma}{\sqrt{3}M_p}}}{3M_p^2(n-1)}
\left(e^{-\frac{\sigma}{\sqrt{3}M_p}}-1\right)^{\frac{1}{n-1}} \nonumber \\
&\qquad
+ \frac{n}{3M_p^2(n-1)^2}
\left(e^{-\frac{\sigma}{\sqrt{3}M_p}}-1\right)^{\frac{2-n}{n-1}}
\Biggr].
\end{align}
One can now look at the critical points and fixed points at infinity for the specific cases of even and odd values of $n$.
\subsubsection{Positive Even $n$ -} \label{sec:poseven}
\begin{itemize}
    \item $\mathbf{n=2}$:\quad In this case, equations (\ref{eq:ffp1}),(\ref{eq:ffp2}) and (\ref{eq:crit1}) show that there exists only one finite fixed point $(H_0,\sigma_0)\equiv(0,0)$ and its attracting basin is the entire allowed phase space. This is the well-known global attractor-to-GR present in Starobinsky model, which has been studied previously. There are no fixed points at infinity as the potential asymptotically saturates to a non-zero finite constant as $\sigma\rightarrow-\infty$, which cannot satisfy equation (\ref{eq:r1}) as $\bar r \rightarrow 1$ \cite{Faraoni:2012bf}.\\
    Since $\sigma_0=0$ is a local minima, existence of the Lyapunov function $L_1$ implies that the fixed point $(0,0)$ is stable.   
    \item $\mathbf{n>2}$: \quad In this case, according to equations (\ref{eq:ffp1}),(\ref{eq:ffp2}),(\ref{eq:crit1}) and (\ref{eq:crit2}) there exists two finite fixed points $P_1:(H_0,\sigma_0)\equiv(0,0)$ and $P_2:(H_0,\sigma_0)\equiv(\sqrt{\frac{V(\sigma_u)}{6M_p^2}},\sigma_u)$. The fixed point $P_1$ continues to be stable in this case as well. However, since $\sigma=\sigma_u$ is a local maxima of the potential, due to the existence of the Lyapunov function $L_2$, the fixed point $P_2$ is unstable. On the other hand, there is also a fixed point at infinity in this case $P_3: (H_0,\sigma_0)\rightarrow(0,-\infty)$ but that corresponds to a future curvature singularity of the $f(R)$ theory because according to the map between $R$ and $\sigma$ given in equation (\ref{eq:RmapSigma}), $\sigma\rightarrow-\infty \implies R\rightarrow\infty$. Therefore, here the attracting basin of the fixed point $P_1$ is not the entire allowed phase space.  
\end{itemize}

\subsubsection{Positive Odd $n$ - Hybrid dynamical system} \label{sec:hybrid}
As mentioned earlier, for positive odd values of $n$, the map $R\mapsto\sigma$ is not one-to-one, therefore when $R$ flows smoothly from positive to negative values, $\sigma$ has to bounce back from $\sigma=0$ in order to be a consistent representation of the Jordan frame system, but the bounce is not provided dynamically by the potential and therefore has to imposed consistently, by hand. We will do so by utilizing the concept of Hybrid dynamical system. It should be noted that a phase analysis in the Jordan frame would not suffer from this issue, however it is interesting that simple techniques from Hybrid dynamical system \cite{e4de97e1c1334dc99ffa1394e35ae6d9} allow us to deal with this issue in the Einstein frame as well. \\
\noindent \textbf{Hybrid Automaton Model:}\quad For this case only, we will change our independent dynamical variables from $(H,\sigma)$ to $(\sigma,\dot\sigma)$. In the hybrid automaton model this system undergoes an autonomous \textbf{jump} in the continuous state $(\sigma,\dot\sigma)$ at an \textbf{impact time} $\tau$, given by the \textbf{impact rule}:
\begin{equation}
    \dot\sigma(\tau^{1\#})=-\dot\sigma(\tau^{0\#})\quad \text{when} \quad \sigma=0\,\,\, \land\,\,\, \dot\sigma\geq 0
\end{equation}
where $\tau^{0\#}$ and $\tau^{1\#}$ are the time instants ``just before" and ``just after" the impact. This impact rule imposes the bouncing condition in the Einstein frame scalar representation. This discrete jump is consistently embedded in the continuous dynamics via the following \textbf{Compatibility conditions}:
\begin{align}
    \lim_{t\uparrow\tau}\sigma(t) &=\sigma(\tau^{0\#}) \qquad\qquad\qquad\qquad \text{(left-hand limit)} \nonumber \\
    &=\sigma(\tau^{1\#})=\lim_{t\downarrow\tau}\sigma(t) \qquad\,\,\,\,\,\,\,\, \text{(right-hand limit)} \nonumber \\
    \lim_{t\uparrow\tau}\dot\sigma(t)&=\dot\sigma(\tau^{0\#}) \nonumber \\
    \lim_{t\downarrow\tau}\dot\sigma(t)&=\dot\sigma(\tau^{1\#})
\end{align}
The complete dynamics of the Hybrid automaton model corresponding the the $f(R)$ theories with positive odd values of $n$ can be written in a more compact way as the \textbf{Event flow formula}:
\begin{equation}
\left|\quad
\begin{aligned}
\sigma \leq 0, &\qquad \ddot{\sigma}=-3H\dot\sigma-V'(\sigma), \\
\sigma=0, &\qquad \dot{\sigma}=0, \\
\sigma=0, &\qquad \dot{\sigma}^{+}=-\dot{\sigma}^{-}.
\end{aligned}
\right.
\end{equation}
The fixed points of this system can be found by the standard methods discussed before and yields two finite fixed points $P_1:(H_0,\sigma_0)\equiv(0,0)$ and $P_2:(H_0,\sigma_0)\equiv\left(\sqrt{\frac{V(\sigma_u)}{6M_p^2}},\sigma_u\right)$. The fixed point $P_2$ is unstable in this case as well (by existence of Lyapunov function $L_2$). There also exists a fixed point at infinity which corresponds to a future-singularity. \\
\noindent Proving the stability of the fixed point $P_1$ needs more care because $\sigma=0$ is precisely where the impact rule is imposed. Since $\sigma^+=-\sigma^- \implies (\sigma^+)^2=(\sigma^-)^2$ the Lyapunov function satisfies the following relation across an impact:
\begin{equation}
    L_1^+=L_1^-
\end{equation}
Away from an impact, the function still satisfies $\dot L_1=-3H\dot\sigma^2$. Therefore, since there exists a function $L_1$ satisfying $\dot L_1\leq 0$ which is non-increasing on the sequence $\tau_1,\tau_2,..$ of event times, the fixed point $(0,0)$ is stable \cite{e4de97e1c1334dc99ffa1394e35ae6d9}. 

\noindent Therefore, for any positive integral value of $n$ in $f(R)=R+\alpha R^n$ gravity, there exists a stable fixed point $(H_0,\sigma_0,\dot\sigma_0)\equiv(0,0,0)$ in the vacuum cosmological dynamics, which corresponds to the theory of gravitation effectively becoming General Relativity because the extra degree of freedom which arises due to the higher curvature modification becomes non-dynamical. However, unlike the case of $n=2$ where the GR fixed point is a global attractor thus making the convergence of the $f(R)$ theory to GR inevitable in the late universe, when $n>2$ the GR fixed point is only a local attractor. Therefore in those cases only in the cosmological evolutions which have no future curvature singularity does the scalar degree of freedom become non-dynamical, thus the theory converging to GR in the late universe.   

\subsection{Negative Power $n=-m$ where $m>0$:} \label{sec:inverse}
\noindent As mentioned earlier, the class of $f(R)$ models with inverse powers of $R$ that we have considered suffer from unstable growth of linearized perturbations in presence of matter (Dolgov-Kawasaki Instability \cite{Dolgov:2003px}). Then why are we interested in such models? It turns out that such models are the high curvature ($R>>\mu^2$) effective descriptions of the viable Hu-Sawicki Gravity \cite{Hu:2007nk} given by $f_{HS}(R)$ in equation (\ref{eq:HS}). Interestingly, in \cite{Hu:2007nk} it was noted that $\mu$ has to be chosen such that $R_0>>\mu^2$ ($R_0$ being the current value of Ricci scalar of the universe), which means that the high-curvature approximation, and therefore the effective description, is valid for the entire expansion history of the universe so far. Therefore, it is in this sense that we are interested in the phase dynamics of such $f(R)$ models with inverse powers of $R$.
\begin{align}
    f_{HS}(R)&=R+\tilde{f}(R)=R-\mu^2\frac{c_1(\frac{R}{\mu^2})^m}{1+c_2(\frac{R}{\mu^2})^m} \label{eq:HS}\\
    &=R + \frac{\alpha}{R^m}+...
\end{align}
where $\alpha=\frac{c_1\mu^{2+2m}}{c_2^2}$. \\
Now that the motivation of considering such models have been laid out, one can proceed with the fixed point analysis by first noting that $f'(R)>0$ fixes the sign of $\alpha$ to be negative. Therefore the first and second derivatives of the potentials can be calculated simply by substituting $n=-m$ and $\alpha=-\tilde\alpha$ into equations (\ref{eq:pospot}) and (\ref{eq:pospot2}),

\begin{equation} \label{eq:bpospot}
    V'(\sigma)=\frac{\tilde\alpha M_p^2(m+1)}{(\tilde\alpha m)^{\frac{m}{m+1}}}\left[ \frac{2e^{\frac{2\sigma}{\sqrt{3}M_p}}}{\sqrt{3}M_p}\left(e^{-\frac{\sigma}{\sqrt{3}M_p}}-1\right)^{\frac{m}{m+1}}-\frac{me^{\frac{\sigma}{\sqrt{3}M_p}}}{\sqrt{3}M_p(m+1)}\left(e^{-\frac{\sigma}{\sqrt{3}M_p}}-1\right)^{-\frac{1}{m+1}} \right]
\end{equation}
\begin{align}\label{eq:bpospot2}
        V''(\sigma)
        &= \frac{\tilde\alpha M_p^2(m+1)}{(\tilde\alpha m)^{\frac{m}{m+1}}}
        \Biggl[
        \frac{4e^{\frac{2\sigma}{\sqrt{3}M_p}}}{3M_p^2}
        \left(e^{-\frac{\sigma}{\sqrt{3}M_p}}-1\right)^{\frac{m}{m+1}}-\frac{me^{\frac{\sigma}{\sqrt{3}M_p}}}{3M_p^2(m+1)}
        \left(e^{-\frac{\sigma}{\sqrt{3}M_p}}-1\right)^{-\frac{1}{m+1}} \nonumber \\
        &\qquad
        - \frac{m}{3M_p^2(m+1)^2}
        \left(e^{-\frac{\sigma}{\sqrt{3}M_p}}-1\right)^{-\frac{2+m}{m+1}}
        \Biggr].
    \end{align}
The first thing to note is that now both the first and second derivatives of the potential contain terms which are singular at $\sigma=0$. Therefore $(0,0)$ is no longer a critical point. Moreover $(0,0)$ corresponds to the future curvature singularity $R\rightarrow\infty$ of the $f(R)$ theory. So the only finite critical point is:

\begin{equation}
    \sigma_0=-\sqrt{3}M_p\ln{\frac{2m+2}{m+2}}\equiv\sigma_u \quad \forall\, m\geq1 \label{eq:bcrit2}
\end{equation}

\noindent However this critical point is a maxima of the potential, which can be checked easily by evaluating $V''(\sigma_u)$ . Therefore by theorem 2 discussed in section [\ref{sec:Lyapunov}], due to the existence of a Lyapunov function $L_2$ this critical point corresponds to an unstable fixed point $(H_0,\sigma_0)\equiv\left(\sqrt{\frac{V(\sigma_u)}{6M_p^2}},\sigma_u\right)$ for all values of $m$. We therefore also don't have to worry about $m$ being odd now because all the interesting dynamics, that is around the stable fixed points, belong to the continuous part of the corresponding Hybrid Automaton model.\\

\noindent On the other hand there is a fixed point at infinity which is of interest for these inverse power models. In the limit $\sigma\rightarrow-\infty$ the potential tends to $0$. Therefore as discussed in section [\ref{sec:infinity}] this satisfies the $\cos\theta=0$ scenario of equations (\ref{eq:r1}) and (\ref{eq:r2}) in the limit $\bar r\rightarrow1$. Therefore $(H,\sigma)\rightarrow(0,-\infty)$ is indeed a fixed point at infinity. In fact, we will now show that the fixed point at $\sigma\to-\infty$ is an
exact power-law attractor of  equations (\ref{eq:hamiltonC})-(\ref{eq:eomSc}), with fixed equation of state. From equation (\ref{eq:Vsigma}), a short calculation gives
the leading behavior of the potential in this limit,
\begin{equation}
V(\sigma)\;\longrightarrow\;V_0\exp\!\left(\lambda\,\frac{\sigma}{M_p}\right),
\qquad
\lambda\equiv\frac{m+2}{\sqrt3\,(m+1)},
\qquad V_0>0, \label{eq:4.24}
\end{equation}
for $n=-m$ ($\tilde\alpha\equiv-\alpha>0$; the same expression follows from
the general-$n$ formula $\lambda=(n-2)/[\sqrt3(n-1)]$ evaluated at
$n=-m$). Note $\lambda^2=(m+2)^2/[3(m+1)^2]\in(1/3,3/4)$ for every $m\geq 1$,
so $\lambda^2<3$ always and $V_0$ is finite and positive.
 
For a pure exponential potential, (\ref{eq:hamiltonC})-(\ref{eq:eomSc}) admit an exact scaling solution of Halliwell type \cite{Halliwell:1986ja,Wetterich:1987fm}:
\begin{equation}
H(t)=\frac{p}{t}, \qquad
\sigma(t)=\sigma_*-\frac{2M_p}{\lambda}\,\ln\!\left(\frac{t}{t_*}\right),
\qquad t>0, \label{eq:e0}
\end{equation}
so that $\dot\sigma=-B/t$ with $B\equiv2M_p/\lambda$. Then $V(\sigma(t))=V_0\exp(\lambda\sigma_*/M_p)\,t_*^{\lambda B/M_p}\,t^{-\lambda B/M_p}\equiv C/t^2$, since $\lambda B/M_p=2$ by construction - the power of $t$ in $V$ matches that of $H^2$ and $\dot\sigma^2$ automatically, which is precisely what makes the exponential potential compatible with a power-law ansatz at all. Substituting into (\ref{eq:eomSc}) fixes $C$ in terms of
$p$,
\begin{equation}
C=\frac{2M_p^2(3p-1)}{\lambda^2}, \label{eq:e1}
\end{equation}
while substituting into the Hamiltonian constraint (\ref{eq:hamiltonC}) fixes $C$ independently as
\begin{equation}
C=6M_p^2p^2-\frac{2M_p^2}{\lambda^2}. \label{eq:e2}
\end{equation}
Equating (\ref{eq:e1}) and (\ref{eq:e2}) gives
\begin{equation}
p=\frac{1}{\lambda^2}, \qquad
C=\frac{2M_p^2(3-\lambda^2)}{\lambda^4}>0 \ \ (\text{since }\lambda^2<3),
\label{eq:e3}
\end{equation}
so that (\ref{eq:e0}) is an exact solution. Using $\rho_\sigma=\tfrac12\dot\sigma^2+V$, $p_\sigma=\tfrac12\dot\sigma^2-V$ (which one can check, directly satisfies $\dot\rho_\sigma+3H(\rho_\sigma+p_\sigma)=0$ by virtue of (\ref{eq:eomSc}), as it must), the scalar field settles onto the equation of state
\begin{equation}
w_\sigma=\frac{p_\sigma}{\rho_\sigma}=\frac{2}{3}\lambda^2-1, \label{eq:4.29}
\end{equation}
Equivalently $w_\sigma=-1+2/(3p)$, the standard power-law relation, provided as a consistency check on (\ref{eq:e3}). Since $\lambda^2\in(1/3,3/4)$ for every $m\geq1$, this attractor is always accelerating ($p>1$): explicitly $p>1 \Leftrightarrow \lambda^2<1 \Leftrightarrow m>(\sqrt3-1)/2\approx0.366$,
which every integer $m\geq1$ studied in this work satisfies. So the late-time endpoint of every inverse-power model in this family is a scaling quintessence attractor observationally similar to Dark Energy with equation of state $w_\sigma$ fixed somewhere in $(-7/9,-1/2)$ depending on $m$. This is also the solution plotted as the ``scaling attractor'' curve in Figure [\ref{fig:panel6}] . The field's kinetic and potential energy scale together in fixed proportion, with $\dot\sigma/H=-2M_p\lambda$ exactly constant along the flow) - distinct from, and not to be confused with, the background-tracking scaling solution of
Copeland-Liddle-Wands type \cite{Copeland:1997et}.

\subsection{Numerical phase plots and trajectories for visualization}

\begin{figure}[!h]
    \centering

    \begin{subfigure}{0.3\linewidth}
        \centering
        \includegraphics[width=\linewidth, height=4.6cm]{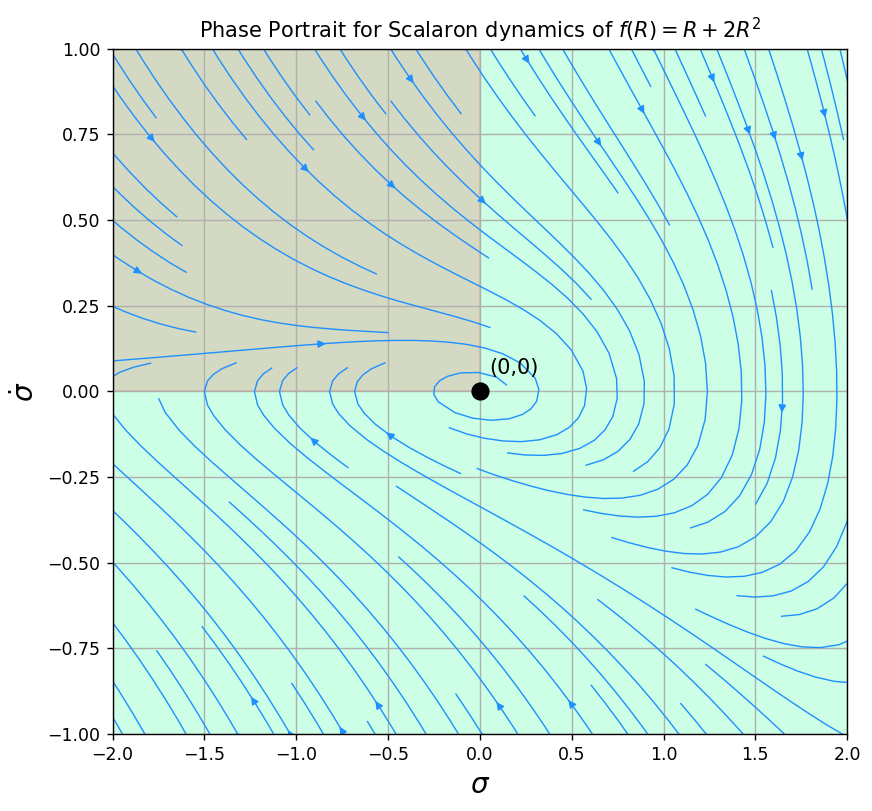} 
        \caption{$n=2$}\label{fig:a}
    \end{subfigure}
    \hspace{0.1cm}
    \begin{subfigure}{0.3\linewidth}
        \centering
        \includegraphics[width=\linewidth]{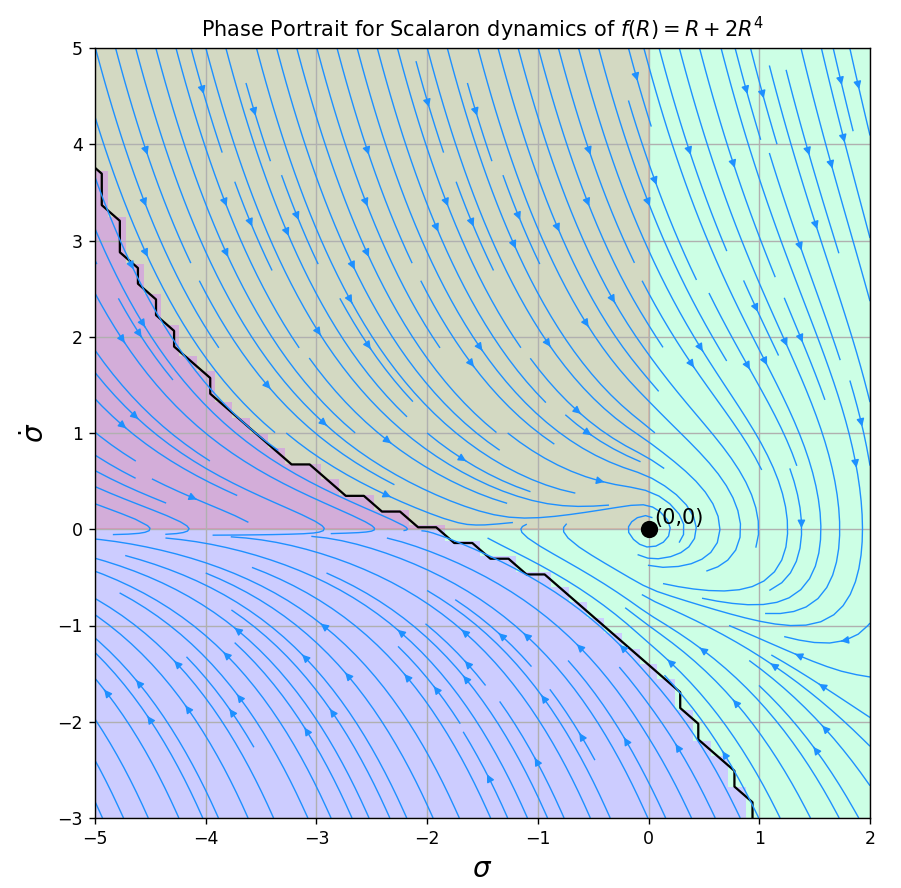}
        \caption{$n=4$}\label{fig:b}
    \end{subfigure}
    \hspace{0.1cm}
    \begin{subfigure}{0.3\linewidth}
        \centering
        \includegraphics[width=\linewidth]{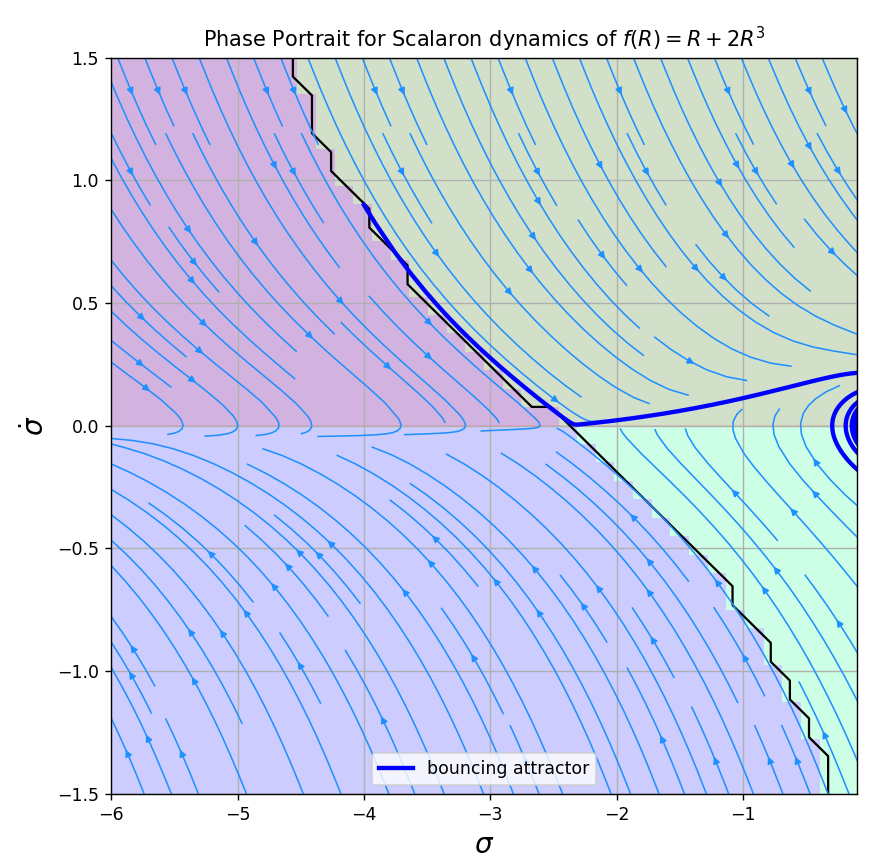}
        \caption{$n=3$}\label{fig:c}
    \end{subfigure}
    \vspace{0.5cm}
    \begin{subfigure}{0.3\linewidth}
        \centering
        \includegraphics[width=\linewidth, height=4cm]{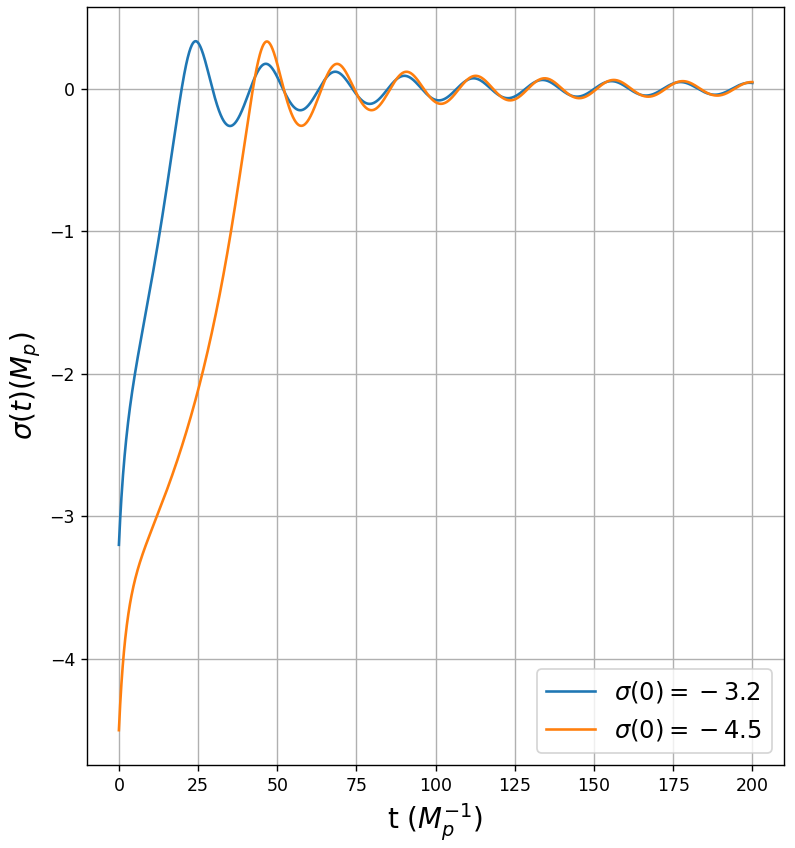} 
        \caption{$n=2$}\label{fig:d}
    \end{subfigure}
    \hspace{0.1cm}
    \begin{subfigure}{0.3\linewidth}
        \centering
        \includegraphics[width=\linewidth, height=4cm]{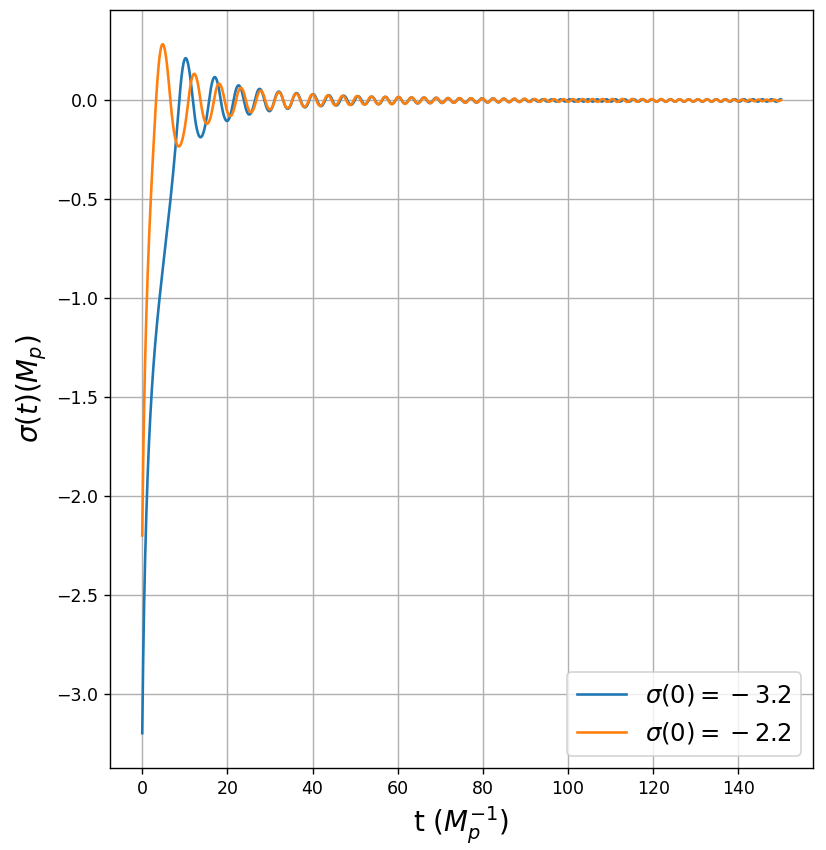}
        \caption{$n=4$}\label{fig:e}
    \end{subfigure}
    \hspace{0.1cm}
    \begin{subfigure}{0.3\linewidth}
        \centering
        \includegraphics[width=\linewidth, height=4cm]{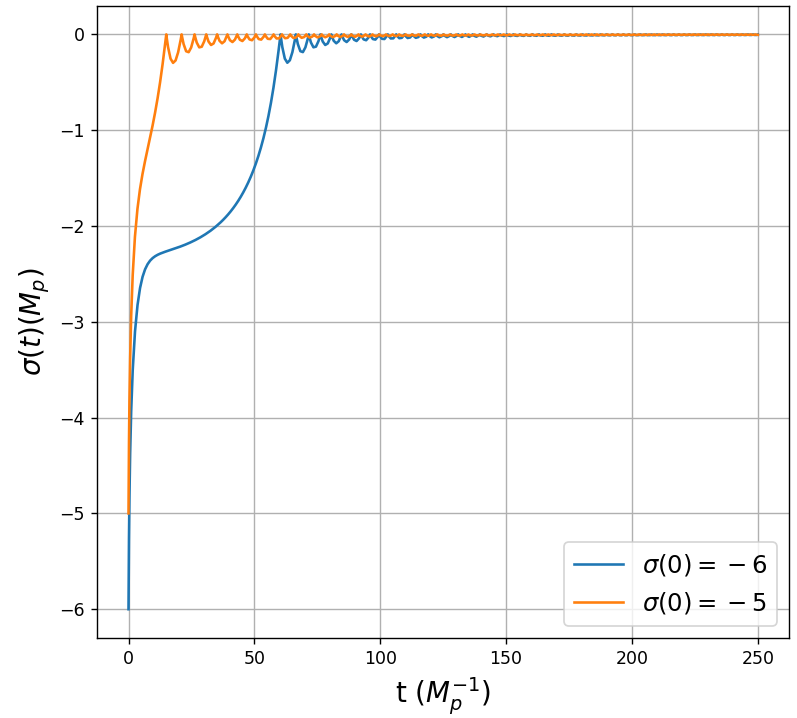}
        \caption{$n=3$}\label{fig:f}
    \end{subfigure}
    \caption{In the upper panel, the convergence basin for the Attractor $(0,0)$ (i.e. GR) and the run-away region are shown in green and blue respectively. The boundary between the two regions denotes the separatrix. The yellow shaded region shows the allowed initial conditions which correspond to the beginning of the universe in the $f(R)$ theory. In the lower panel time evolution of particular initial conditions of the scalar degree of freedom is shown.}
    \label{fig:panel2}
\end{figure}

\begin{figure}[!h]
    \centering

    \begin{subfigure}{0.3\linewidth}
        \centering
        \includegraphics[width=\linewidth]{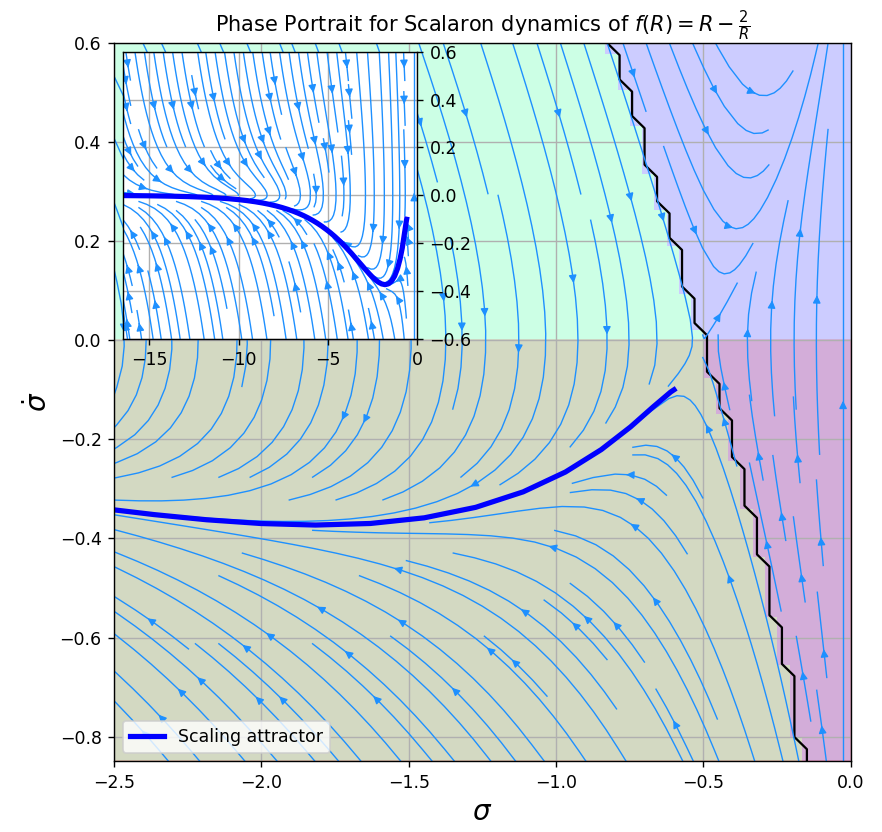} 
        \caption{$m=1$}\label{fig:a}
    \end{subfigure}
    \hspace{1.5cm}
    \begin{subfigure}{0.3\linewidth}
        \centering
        \includegraphics[width=\linewidth]{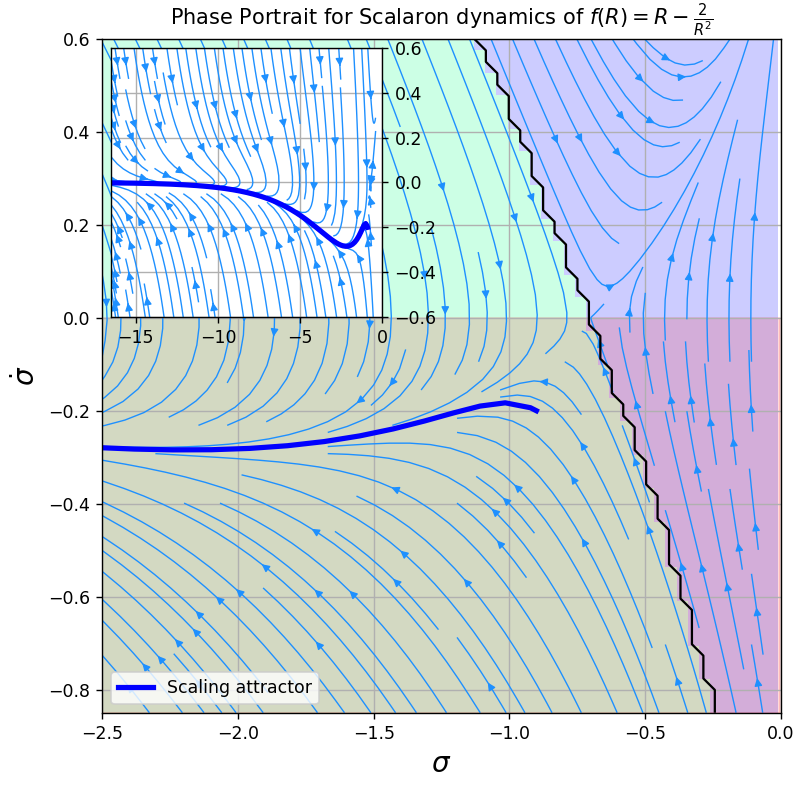}
        \caption{$m=2$}\label{fig:b}
    \end{subfigure}
    \vspace{0.5cm}
    \begin{subfigure}{0.3\linewidth}
        \centering
        \includegraphics[width=\linewidth, height=4.5cm]{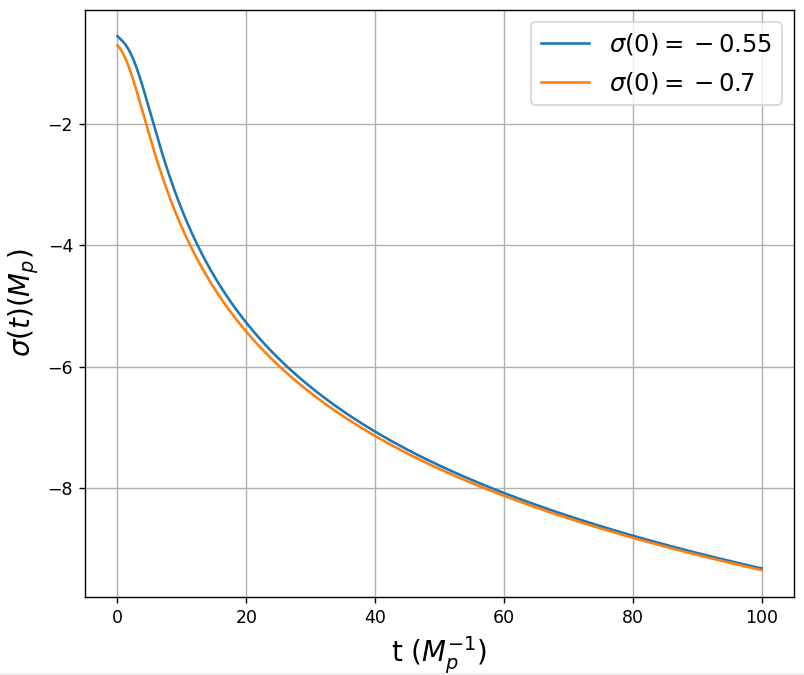} 
        \caption{$m=1$}\label{fig:c}
    \end{subfigure}
    \hspace{1.5cm}
    \begin{subfigure}{0.3\linewidth}
        \centering
        \includegraphics[width=\linewidth, height=4.5cm]{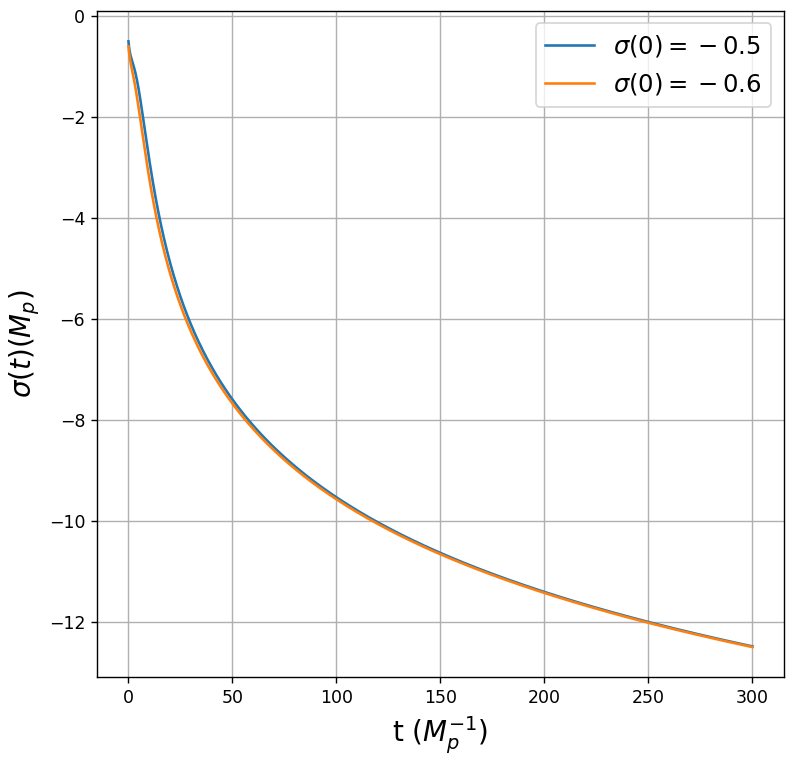}
        \caption{$m=2$}\label{fig:d}
    \end{subfigure}
    \caption{In the upper panel, the convergence basin for the asymptotic attractor and the run-away region are shown in green and blue respectively. The boundary between the two regions denotes the separatrix as before and the yellow shaded region shows the allowed initial conditions. Time evolution shows the saturation/freezing of the scalar degree of freedom in late times.}
    \label{fig:panel6}
\end{figure}

In this subsection we will show numerical phase plots and time evolution trajectories generated by solving the system of equations (\ref{eq:hamiltonC}) and (\ref{eq:eomSc}). Rather than plotting the phase diagrams with $(H,\sigma)$ as the dynamical variables, the variables $(\sigma,\dot\sigma)$ is more intuitive and we know from our discussions in subsection [\ref{sec:psg}] that they are equivalent because of the Hamiltonian constraint. Figures [\ref{fig:panel2}] and [\ref{fig:panel6}] show the phase diagrams for the positive and inverse power $f(R)$ models respectively and also the corresponding time evolution trajectories of the scalar field for particular initial conditions. 

\clearpage

\section{Evolution in presence of Matter -}\label{sec:matter}
All of the above analysis previously was done in absence of matter fields or in a regime where the universe is scalar field dominated. In this section, the evolution of these models in the presence of radiation ($\rho=3p$) and dust ($p=0$) type of matter has been discussed. When a matter action $S_m[\psi]$ is added to the $f(R)$ action (\ref{eq:fRaction}), conversion to the Einstein frame results in the matter becoming non-minimally coupled as $S_m[e^{\frac{\sigma}{\sqrt{3}M_p}}g_{\mu\nu},\psi]$. We extend the vacuum phase space $(\sigma,\dot\sigma)$ of Section [\ref{sec:dynam}] by the
fluid energy density $\rho$, so that $(\sigma,\dot\sigma,\rho)$ are treated as the dynamical variables, with $H$ fixed implicitly through the Hamiltonian constraint. For a barotropic fluid with equation of state $p=w\rho$ (so $w=1/3$ for radiation, $w=0$ for dust), the Friedman equation, scalar field equation of motion and matter equations can be written for general $w$ as
\begin{align}
3H^2 &= \frac{1}{2M_p^2}\left(\frac12\dot\sigma^2+V(\sigma)+\rho\right), \label{eq:frid}\\
\ddot\sigma+3H\dot\sigma+V'(\sigma) &= -\frac{(1-3w)}{2\sqrt3\,M_p}\,\rho, \label{eq:scalar}\\
\dot\rho+3H(1+w)\rho &= \frac{(1-3w)}{2\sqrt3\,M_p}\,\dot\sigma\,\rho, \label{eq:matter}
\end{align}
where $\rho$ and $p$ are the Einstein-frame density and pressure of the fluid defined in terms of the corresponding Jordan-frame quantities $\rho_J$ and $p_J$ as: $\rho=e^{\frac{\sigma}{\sqrt{3}M_p}}\rho_J$ and $p=e^{\frac{\sigma}{\sqrt{3}M_p}}p_J$.
Equation (\ref{eq:matter}) shows immediately that $\rho\equiv0$ is a
dynamically invariant subset of the extended phase space: any
trajectory that starts with $\rho=0$ remains there for all time, and that is exactly the vacuum system of Section [\ref{sec:dynam}]. Moreover from equations (\ref{eq:frid}), (\ref{eq:scalar}) and (\ref{eq:matter}) note that for radiation ($p=\frac{1}{3}\rho$), the energy-density $\rho$ falls as $a^{-4}$ and contributes to the scalar field's equation of motion only via Hubble friction. Therefore, radiation only modifies the evolution of the scalar field transiently; however, as $\rho$ falls and becomes sub-dominant, vacuum dynamics is recovered asymptotically, thereby restoring the attractor-to-GR mechanism discussed in section [\ref{sec:dynam}]. Nevertheless, there could be additional fixed points in the transient dynamics that one should be careful about.

\subsection{Fixed points of the matter-coupled system}

A fixed point of the system of equations (\ref{eq:frid}), (\ref{eq:scalar}) and (\ref{eq:matter}), by definition has to be of the form $(\sigma_0,0,\rho_0)$ and requires $\dot\sigma=0$ together with
\begin{align}
& \dot\sigma_0=0 \\
& V'(\sigma_0) = -\frac{(1-3w)}{2\sqrt3\,M_p}\,\rho_0, \label{eq:slope}\\
& 3H_0(1+w)\rho_0 = 0. \label{eq:h0}
\end{align}
Since $w\neq-1$ here, equation (\ref{eq:h0}) forces $H_0=0$ or $\rho_0=0$.

\begin{itemize}
\item\textbf{Case A} ($\rho_0=0$): Equation (\ref{eq:slope}) reduces to $V'(\sigma_0)=0$, which is exactly the vacuum critical-point condition (\ref{eq:ffp2}). Every fixed point found in Section [\ref{sec:dynam}] therefore turns out to be a fixed point
$(\sigma_0,0,0)$ of the matter-coupled system.

\item\textbf{Case B} ($H_0=0$): The Hamiltonian constraint (\ref{eq:frid}) at $\dot\sigma_0=0$, $H_0=0$ gives $V(\sigma_0)=-\rho_0$. Since $\rho_0\geq0$
for physical matter and $V(\sigma)\geq0$ for every model considered in this work (Figure [\ref{fig:panel1}]), this forces $\rho_0=0$ and $V(\sigma_0)=0$ - which is already contained in Case A.
\end{itemize}

\noindent Hence the matter-coupled system has no finite fixed points beyond the vacuum ones of Section [\ref{sec:dynam}]; all of them lie on the invariant plane $\rho=0$. This holds for both radiation and dust, and for every $n$
(or $m$) considered in this paper. As for the fixed points at infinity, since $\rho\to 0$ and $\dot\sigma\to 0$ asymptotically, and the scalar-matter coupling is of the form $\dot\sigma\rho$ (look at equation \ref{eq:matter}) the system of equations reduce to the vacuum case asymptotically, therefore keeping the fixed point at infinity undisturbed.

\subsection{Stability: an extended Lyapunov function}

For any vacuum fixed point $(\sigma_0,0,0)$ with $V''(\sigma_0)>0$ - i.e.\ those covered by Theorem 1 of Section [\ref{sec:dynam}], with Lyapunov function $L_1$, define
\begin{equation}
\widetilde L_1(\sigma,\dot\sigma,\rho) \;:=\; \frac12\dot\sigma^2+V(\sigma)-V(\sigma_0)+\rho
\;=\; L_1(\sigma,\dot\sigma) + \rho. \label{eq:5.6}
\end{equation}
Since $\rho\geq0$, $\widetilde L_1$ inherits positivity from $L_1$ in a punctured neighborhood (everywhere in a domain $\mathcal{D}$ except at the fixed point) of $(\sigma_0,0,0)$, with $\widetilde L_1(\sigma_0,0,0)=0$. Differentiating $\widetilde L_1$, and using $\dot L_1=-3H\dot\sigma^2-
\frac{(1-3w)}{2\sqrt3M_p}\dot\sigma\rho$,
\begin{align}
\dot{\widetilde L}_1 &= \dot L_1+\dot\rho \notag \\
&= -3H\dot\sigma^2-\frac{(1-3w)}{2\sqrt3M_p}\dot\sigma\rho
\;-\;3H(1+w)\rho+\frac{(1-3w)}{2\sqrt3M_p}\dot\sigma\rho \notag\\
&= -3H\Big[\dot\sigma^2+(1+w)\rho\Big]. \label{eq:liap1b}
\end{align}
Therefore for the expanding branch $H>0$, with $\rho\geq0$ and $w>-1$, the bracket in equation (\ref{eq:liap1b}) is non-negative, so $\dot{\widetilde L}_1\leq0$, vanishing only at $(\sigma_0,0,0)$, thereby proving stability of such a fixed point. Every vacuum-stable fixed point therefore remains stable once radiation or dust is switched on, and the same argument forces $\rho\to0$ asymptotically for both the global attractor at $n=2$ and for the locally-stable $P_1$ at $n>2$. For the unstable point $P_2=(H_0,\sigma_u)$ (where $V''(\sigma_u)<0$), the analogous combination $\widetilde L_2 := L_2-\rho$ gives $\dot{\widetilde L}_2 = 3H\big[\dot\sigma^2+(1+w)\rho\big]\geq0$ by the same cancellation, confirming $P_2$ remains unstable with matter present.

\noindent As for the hybrid (odd $n$) case of Section~4.4.2, note that the impact rule only flips $\dot\sigma\to-\dot\sigma$ and leaves $\rho$ continuous, so $\widetilde L_1^+=\widetilde L_1^-$ across an impact exactly as $L_1^+=L_1^-$ did; the stability argument therefore carries over unchanged.

\begin{figure}[H]
    \centering

    \begin{subfigure}{0.3\linewidth}
        \centering
        \includegraphics[width=0.8\linewidth, height=3cm]{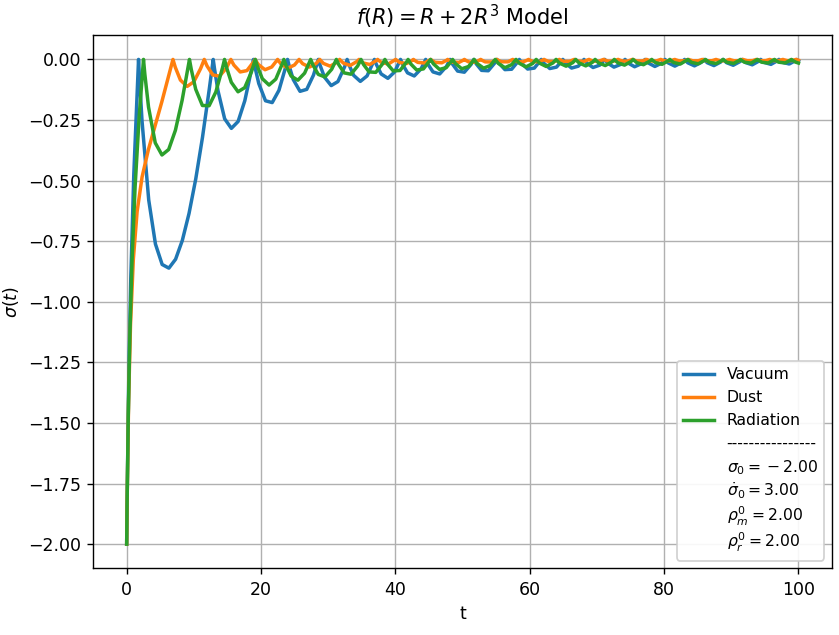} 
        \caption{$n=3$}\label{fig:a}
    \end{subfigure}
    \hfill
    \begin{subfigure}{0.3\linewidth}
        \centering
        \includegraphics[width=0.8\linewidth, height=3cm]{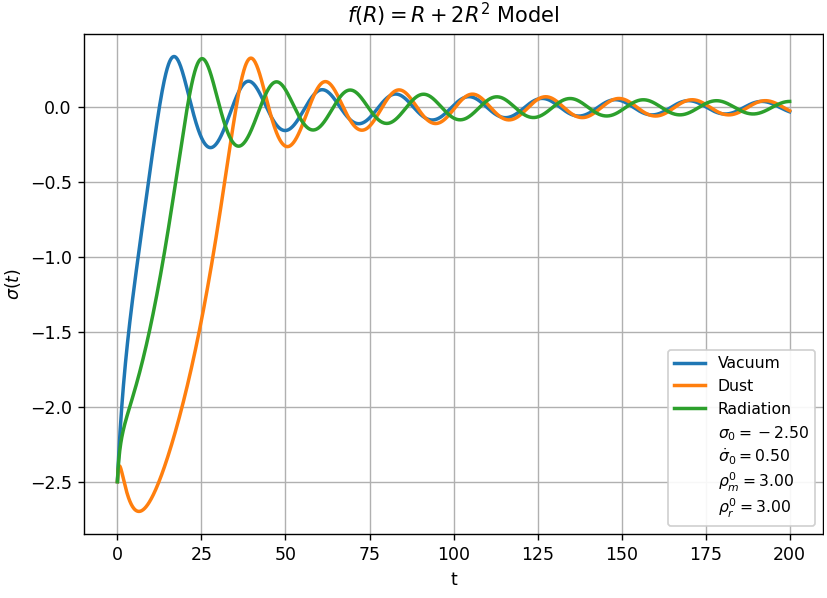}
        \caption{$n=2$}\label{fig:p}
    \end{subfigure}
    \hfill
    \begin{subfigure}{0.3\linewidth}
        \centering
        \includegraphics[width=0.8\linewidth, height=3cm]{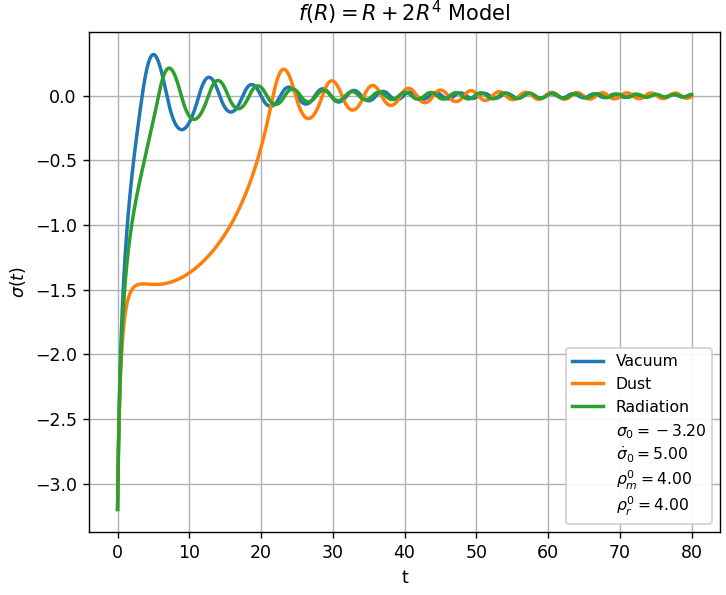}
        \caption{$n=4$}\label{fig:c}
    \end{subfigure}

    \vspace{0.1cm}

    \begin{subfigure}{0.3\linewidth}
        \centering
        \includegraphics[width=0.8\linewidth, height=3cm]{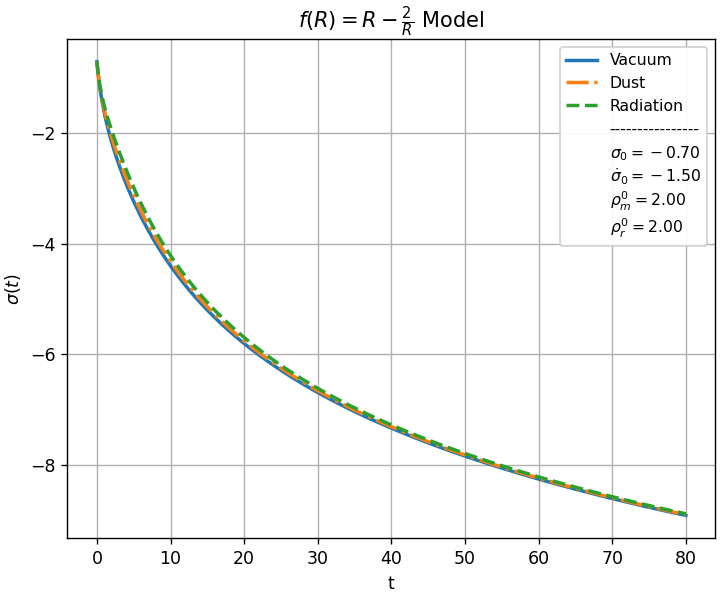}
        \caption{$m=1$}\label{fig:d}
    \end{subfigure}
    \hspace{0.1cm}
    \begin{subfigure}{0.3\linewidth}
        \centering
        \includegraphics[width=0.8\linewidth, height=3cm]{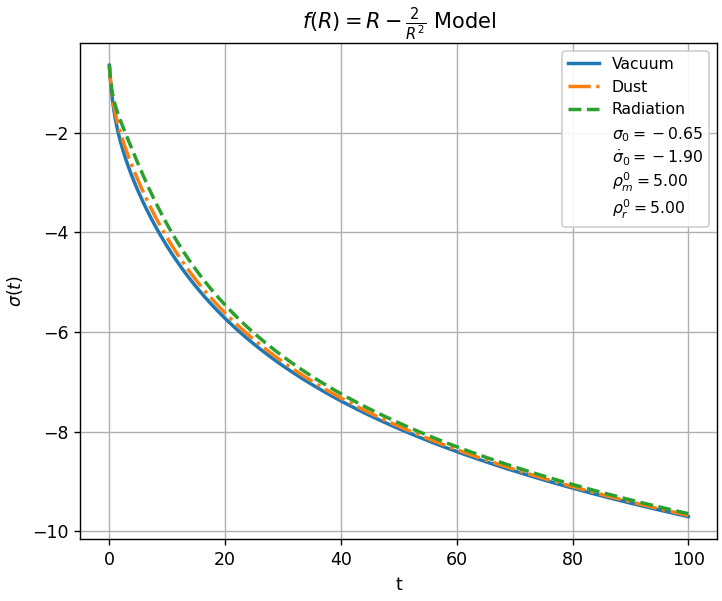}
        \caption{$m=2$}\label{fig:e}
    \end{subfigure}

    \caption{Comparison of scalar-field evolution trajectories in vacuum, radiation and dust starting from same initial conditions is shown here. In these plots $\sigma,\dot{\sigma},\rho_r,\rho_m,t$ are in units of appropriate powers of $M_p$.}
    \label{fig:panel8}
\end{figure}

\noindent Figure [\ref{fig:panel8}] compares the numerically obtained time evolution trajectories of the scalar degree of freedom in vacuum, radiation and dust backgrounds, for a few particular models (given by $n$) and choice of initial conditions. The asymptotic restoration of the vacuum dynamics is clearly visible as predicted by the phase space dynamical system analysis.

\noindent From Figures \ref{fig:panel8}(\subref{fig:a}), \ref{fig:panel8}(\subref{fig:p}) and \ref{fig:panel8}(\subref{fig:c})  we can note that the presence of radiation and dust delays the fall of the scalar field towards $\sigma=0$ and then also slows down the decay of the oscillations. But after some time, the three trajectories merge, therefore restoring vacuum dynamics as expected.

\noindent From Figures \ref{fig:panel8}(\subref{fig:d}) and \ref{fig:panel8}(\subref{fig:e}) we can see that the presence of dust or radiation initially slows the descent of the scalar field towards larger negative values but at late times, both trajectories merge with the vacuum trajectory (blue), as discussed in earlier sections.

\section{Observational Constraints on the Model Parameters -}\label{sec:observations}
As shown in Sections \ref{sec:setup}-\ref{sec:matter}, convergence to a stable fixed point holds for every integer $n$ (or $m$) and every $\alpha$; Appendix A shows $\alpha$ only rescales the potential and the convergence timescale (Figure [\ref{fig:panel4}]), leaving the fixed-point locations ($\sigma_0=0$, $\sigma_u$) and $w_\sigma$ in equation (\ref{eq:4.29}) untouched. This separation - $n$ (or $m$) fixes the structure, $\alpha$ only the rate -- lets us compare each parameter against the literature separately.
 
\noindent For the positive-power branch, \cite{Odintsov:2025eiv} finds $1.962<n<1.978$ or $n=2$ from PLANCK-2018/ACT, and \cite{Saini:2023uqc} gives $\alpha\sim10^{4.48}$ from PLANCK/BICEP3. This window brackets $n=2$ - the one case in our family where $P_1=(0,0)$ is a global attractor rather than only locally stable with a separatrix (Sections [\ref{sec:poseven}]-[\ref{sec:hybrid}]) - while lying outside the integer-$n$ family itself, so it mildly disfavors the $n=3,4,\dots$ regime where the basin structure actually matters. By the separation above, the $\alpha$ bound only fixes how fast inflation ends, not whether $P_1$ attracts.
 
\noindent For the inverse-power branch, \cite{Martinelli:2009ek} constrains $n$ (their notation for our $m$) and $\tilde f_{R0}$ from SNIa, BAO, and Hubble-expansion/age data for different values of the effective matter content $\tilde\Omega_m$ - higher $\tilde\Omega_m$ prefers higher $n$, but also loosens the bound - with $|\alpha|\sim10^{-121}M_p^{2n+2}$ forced separately by $R_0\gg \mu^2$; \cite{Kou:2023gyc, Kumar:2025mzo,Yan:2024jwz} improve this further. None of these bounds engage $\sigma_u$ (\ref{eq:bcrit2}) or $w_\sigma=\tfrac23\lambda^2-1$ (\ref{eq:4.29}), both fixed by $m$ alone: $w_\sigma=-1/2$ ($m=1$), $\approx-0.60$ ($m=2$), $\to-7/9$ as $m\to\infty$ - a discrete prediction no $\alpha$ within the bound can shift. This should be compared to dark-energy data with caution, since the $R\gg\mu^2$ truncation used here is least trustworthy exactly where the attractor lives, $\sigma\to-\infty$ ($R\to0$); a proper comparison needs the un-truncated $f_{HS}(R)$. (As a naming note, the exponent we call $m$ is usually called $n$ in the Hu-Sawicki literature, including \cite{Martinelli:2009ek}; we keep our own notation to avoid a clash with the positive-power $n$.)
 
\noindent Since convergence itself is independent of $\alpha$ and of which integer $n$ (or $m$) is chosen - only its qualitative character (global/local, exact GR/quintessence tail) depends on the choice - these constraints narrow down which member of the family is realized, without threatening the mechanism itself.

\section{Conclusion}\label{sec:conclusion}
This work has given a unified phase-space treatment of the scalar degree of freedom of $f(R)=R+\alpha R^n$ gravity in Einstein-frame representation across every integer $n$, positive and negative alike, combining finite fixed points, Poincar\'e-compactified fixed points at infinity, and non-perturbative Lyapunov stability into a single framework, with analytic results checked against direct numerical evolution. This required one genuinely new technical idea: for every odd $n$, the Einstein-frame potential does not by itself enforce the ``bounce'' at $\sigma=0$ implicit in the $R\leftrightarrow\sigma$ correspondence, and we resolved this by formulating the odd $n$ sector as a hybrid dynamical system with an explicit impact rule (Section [\ref{sec:hybrid}]) - to our knowledge the first use of this framework in this setting.
 
\noindent The attractor-to-GR mechanism itself has been established in various scalar-tensor settings before \cite{Jarv:2010zc,Jarv:2010xm,Jarv:2015kga,Jarv:2011sm,Jarv:2015odu}. What this work adds is a structural account of \textit{how} that convergence happens across the entire $R^n$ family. $n=2$ is singled out as the unique case with a genuinely global attractor; every $n>2$ has only a local one, bounded by an explicit separatrix beyond which trajectories run instead to a curvature singularity (Section [\ref{sec:positive}]). For inverse powers ($n=-m$), the only finite fixed point is unstable, and the true late-time endpoint is the fixed point at infinity - which is approached by a power-law scaling solution of the cosmological equations (Section [\ref{sec:inverse}]), with accelerated expansion and an equation of state $w_\sigma=\tfrac23\lambda^2-1$ fixed entirely by the integer $m$: $-1/2$ at $m=1$, decreasing toward $-7/9$ as $m\to\infty$.
 
\noindent Extending the analysis to radiation and dust (Section [\ref{sec:matter}]), we showed - via an extended Lyapunov function whose construction cancels the matter-scalar coupling term exactly - that every vacuum-stable fixed point remains stable once matter is included, that no fixed point exists anywhere in the extended phase space beyond the vacuum ones. Section [\ref{sec:observations}] showed that current inflationary bounds on $n$ sit almost exactly on the boundary between the global- and local-attractor regimes identified here, while existing bounds on $\alpha$ constrain only the rate of convergence and not the qualitative outcome. Several directions follow naturally. The freeze-out established here is a statement resting on homogeneity and isotropy assumptions; whether the same mechanism persists in anisotropic or inhomogeneous settings remains open, as does the corresponding local, perturbative stability question - the Dolgov-Kawasaki instability that excludes the inverse-power branch on solar-system scales is a statement about $\delta R$, and it would be worth understanding to what extent it is the perturbative shadow of the very fixed-point structure derived here, rather than a logically separate constraint. On the model side, this study treated positive and negative power corrections separately; the phase space of $f(R)$ actions containing both simultaneously is a natural and more phenomenologically realistic next step. Finally, since the discrete $w_\sigma(m)$ predicted here follows from the truncated $R+\alpha R^{-m}$ action rather than the exact Hu-Sawicki form it approximates, confronting it with dark energy data will first require repeating this analysis for the un-truncated $f_{HS}(R)$, for which the $\sigma\to-\infty$ limit explored here need not coincide with the $R\gg\mu^2$ regime in which the truncation is justified.

\noindent The class of f(R) models considered in this work is among the simplest members of a much larger space of possible corrections to the Einstein-Hilbert action - simpler, for instance, than logarithmic or exponential corrections - and as a truncated, leading-order term in that wider space it should not be expected to provide an accurate description of physical cosmology on its own. The merit of studying such models lies instead in the leading-order understanding they provide of two structurally distinct classes of modification over the dynamical phase space of General Relativity, which can guide the search for better-motivated physical descriptions of the universe we live in. One would then also like to ask whether these corrections over GR are classical or truly quantum in nature - potentially requiring a re-interpretation of the attractor-to-GR mechanism in the light of a quantum origin for the effective theory. This, in turn, would mean checking, model by model, whether it survives once the full tower of corrections is kept rather than a single truncated term - a question this paper does not attempt to answer in general, but one the phase-space language developed here seems well suited to eventually address.

\acknowledgments
GC is supported by Indian Association for the Cultivation of Science (IACS) Masters Fellowship. He also thanks Soham Bhattacharya for many useful discussions.

\bibliographystyle{unsrt}
\bibliography{main.bib}

\appendix
\section{Appendix I: Qualitative features of varying $\alpha$ in the $f(R)$ action}

\begin{figure}[H]
    \centering

    \begin{subfigure}{0.45\linewidth}
        \centering
        \includegraphics[width=0.8\linewidth]{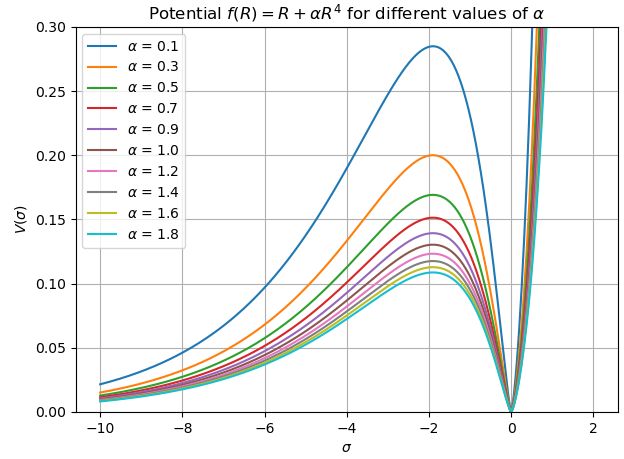} 
        \caption{Varying $\alpha$ for $n=4$}\label{fig:a}
    \end{subfigure}
    \hspace{0.3cm}
    \begin{subfigure}{0.45\linewidth}
        \centering
        \includegraphics[width=0.8\linewidth]{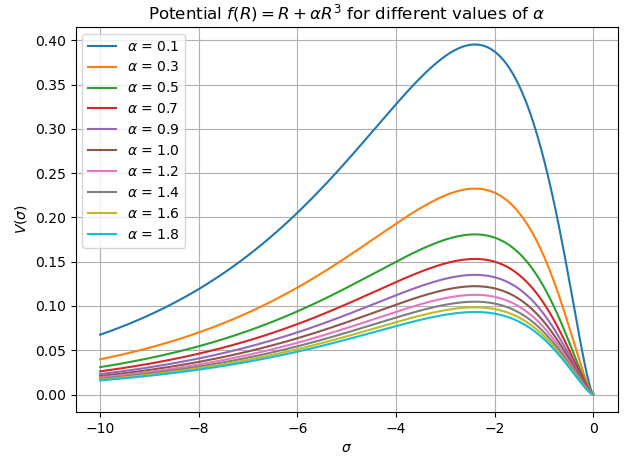}
        \caption{Varying $\alpha$ for $n=3$}\label{fig:b}
    \end{subfigure}

    \vspace{0.1cm}

    \begin{subfigure}{0.45\linewidth}
        \centering
        \includegraphics[width=0.8\linewidth]{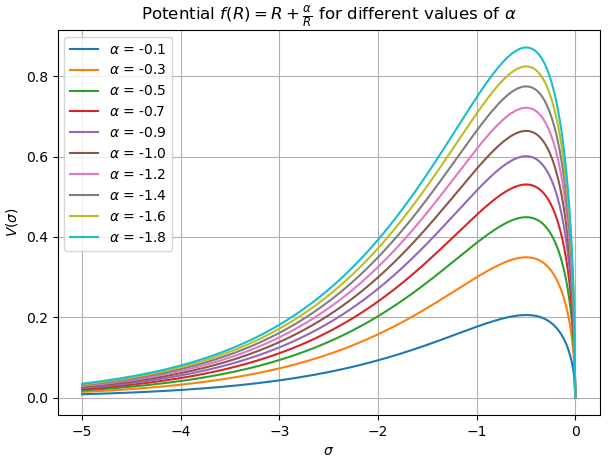} 
        \caption{Varying $\alpha$ for $n=-m=-1$}\label{fig:c}
    \end{subfigure}
    \hspace{0.3cm}
    \begin{subfigure}{0.45\linewidth}
        \centering
        \includegraphics[width=0.8\linewidth]{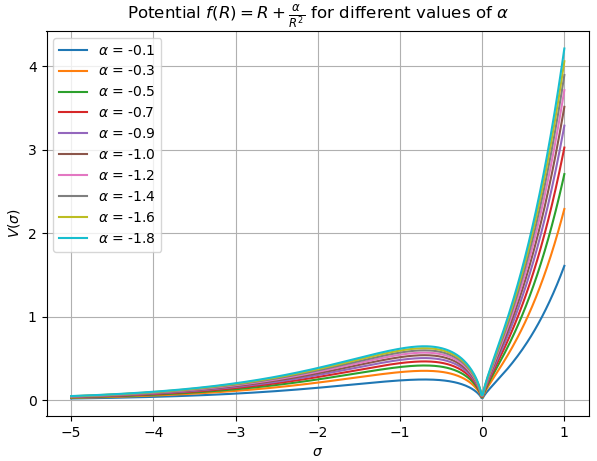}
        \caption{Varying $\alpha$ for $n=-m=-2$}\label{fig:d}
    \end{subfigure}

    \caption{Forms of the potentials for various choices of $\alpha$ ($\sigma$ is in units of $M_p$).}
    \label{fig:panel5}
\end{figure}

The shape of the Einstein frame potentials only experiences a constant scaling when $\alpha$, the coefficient of the $f(R)$ corrections is changed. The scaling is apparent from the form of the potentials but is also shown in figures [\ref{fig:panel5}].

\begin{figure}[!h]
    \centering

    \begin{subfigure}{0.3\linewidth}
        \centering
        \includegraphics[width=\linewidth, height=4cm]{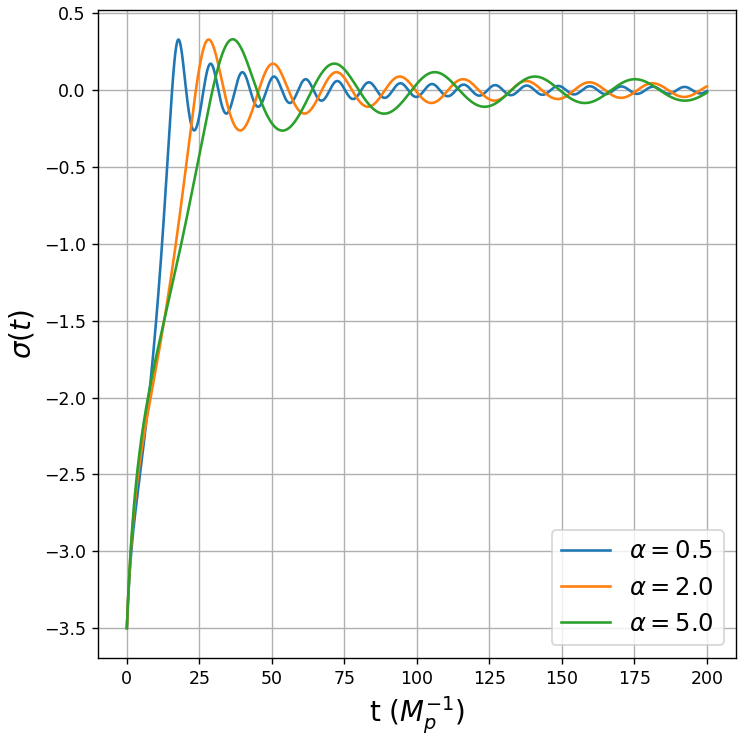} 
        \caption{$n=2$}\label{fig:a}
    \end{subfigure}
    \hspace{0.1cm}
    \begin{subfigure}{0.3\linewidth}
        \centering
        \includegraphics[width=\linewidth, height=4cm]{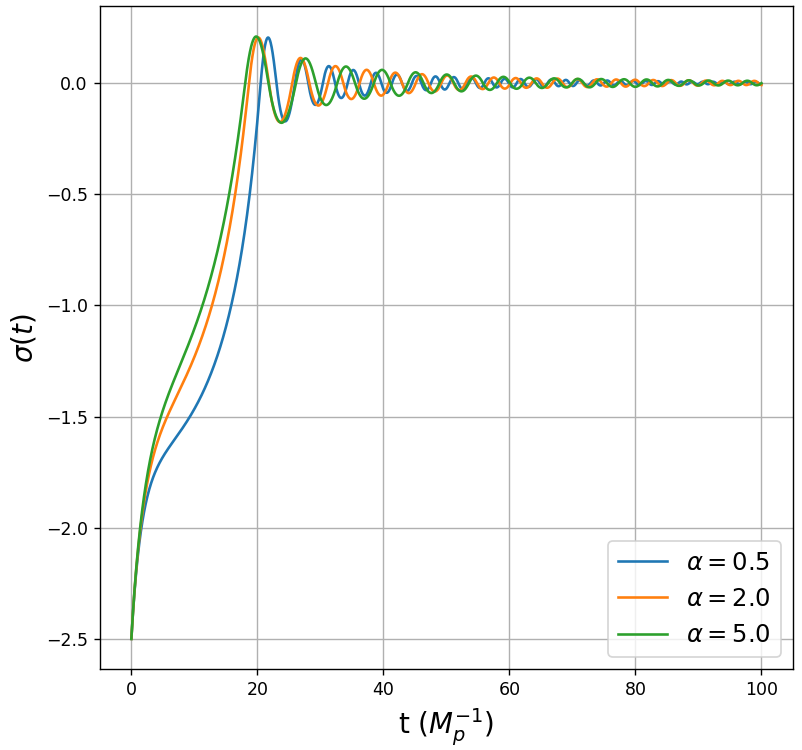}
        \caption{$n=4$}\label{fig:b}
    \end{subfigure}
    \hspace{0.1cm}
    \begin{subfigure}{0.3\linewidth}
        \centering
        \includegraphics[width=\linewidth, height=4cm]{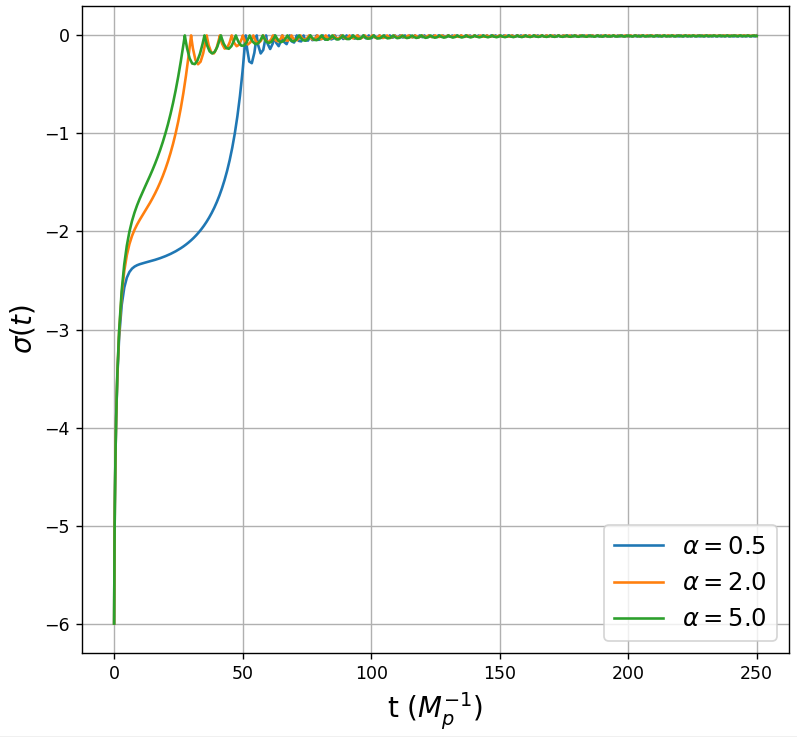}
        \caption{$n=3$}\label{fig:c}
    \end{subfigure}
    \caption{Effect of varying $\alpha$ on the time evolution of the scalar degree of freedom}
    \label{fig:panel4}
\end{figure}

\noindent Although not specifically relevant, Figures \ref{fig:panel4}(\subref{fig:a}), \ref{fig:panel4}(\subref{fig:b}) and \ref{fig:panel4}(\subref{fig:c}) show that increasing $\alpha$ results in a decrease in the oscillation frequency of the scalar field near $\sigma=0$. These figures also highlight another way in which the $n=2$ case is special among all positive integral powers: note that for both $n=3$ and $n=4$, a higher value of $\alpha$ makes $\sigma$ fall towards $0$ earlier, but for the case of $n=2$, a higher value of $\alpha$ causes a delay in the fall of the scalar field towards $\sigma=0$. 

\end{document}